\documentclass[useAMS,usenatbib]{mnras}
\usepackage{amsmath}
\usepackage{amssymb}
\usepackage{graphicx}
\usepackage{lscape}
\usepackage{listings}
\usepackage{hyperref}
\usepackage{amstext} 
\usepackage{array}   
\newcolumntype{L}{>{$}l<{$}} 

\newcommand{\Lsun}{\mbox{$L_{\odot}$}} 
 
\newcommand{\Rjup}{\mbox{$R_{\rm Jup}$}} 
\newcommand{\Mjup}{\mbox{$M_{\rm Jup}$}}
\newcommand{\likL}{\mathcal{L}}
\newcommand{\ltsimeq}{\raisebox{-0.6ex}{$\,\stackrel
        {\raisebox{-.2ex}{$\textstyle <$}}{\sim}\,$}}
\newcommand{\gtsimeq}{\raisebox{-0.6ex}{$\,\stackrel
        {\raisebox{-.2ex}{$\textstyle >$}}{\sim}\,$}}
\newcommand{\chhhh}{\mbox{${\rm CH_4}$ }}
\newcommand{\hho}{\mbox{${\rm H_{2}O}$ }} 
\newcommand{\coo}{\mbox{${\rm CO_{2}}$ }} 

\newcommand{\teff}{\mbox{$T_{\rm eff}$ }}
\title[Cloud busting]{Cloud busting: enstatite and quartz clouds in the atmosphere of 2M2224-0158}
\author[Burningham et al]{Ben Burningham$^{1}$\thanks{E-mail:
    B.Burningham@herts.ac.uk}, Jacqueline K. Faherty$^{2}$, Eileen C. Gonzales$^{1,2,3,4,5}$\thanks{51 Pegasi b Fellow}, Mark S. Marley$^{6}$,
    \newauthor  Channon Visscher$^{7,8}$, Roxana Lupu$^{6,9}$, Josefine Gaarn$^{1}$, Michelle Fabienne Bieger$^{1,10}$, 
    \newauthor Richard Freedman$^{6,11}$, Didier Saumon$^{12}$, \\ 
$^{1}$ Centre for Astrophysics Research, Department of Physics, Astronomy and Mathematics, University of Hertfordshire, Hatfield AL10 9AB \\
$^{2}$ Department of Astrophysics, American Museum of Natural History, New York, NY 10024, USA\\
$^{3}$ Department of Astronomy and Carl Sagan Institute, Cornell University, 122 Sciences Drive, Ithaca, NY 14853, USA \\
$^{4}$ The Graduate Center, City University of New York, New York, NY 10016, USA \\
$^{5}$ Department of Physics and Astronomy, Hunter College, City University of New York, New York, NY 10065, USA \\ 
$^{6}$ NASA Ames Research Center, Mail Stop 245-3, Moffett Field, CA 94035, USA \\
$^{7}$ Department of Chemistry, Dordt University, Sioux Center, IA 51250, USA \\
$^{8}$ Space Science Institute, Boulder, CO 80301, USA \\
$^{9}$ Bay Area Environmental Research Institute, 625 2nd Street, Suite 209, Petaluma, CA 94952, USA\\
$^{10}$ College of Engineering, Mathematics and Physical Sciences,  University of Exeter, North Park Road, Exeter, United Kingdom \\
$^{11}$ SETI Institute, Mountain View, CA 94043, USA\\
$^{12}$ Los Alamos National Laboratory, P.O. Box 1663, MS F663, Los Alamos, NM 87545, USA \\
}

\begin{document}
%
%
%
%


\def\aj{\rm{AJ}}                   
\def\araa{\rm{ARA\&A}}             
\def\apj{\rm{ApJ}}                 
\def\apjl{\rm{ApJ}}                
\def\apjs{\rm{ApJS}}               
\def\ao{\rm{Appl.~Opt.}}           
\def\apss{\rm{Ap\&SS}}             
\def\aap{\rm{A\&A}}                
\def\aapr{\rm{A\&A~Rev.}}          
\def\aaps{\rm{A\&AS}}              
\def\azh{\rm{AZh}}                 
\def\baas{\rm{BAAS}}               
\def\jrasc{\rm{JRASC}}             
\def\memras{\rm{MmRAS}}            
\def\mnras{\rm{MNRAS}}             
\def\pra{\rm{Phys.~Rev.~A}}        
\def\prb{\rm{Phys.~Rev.~B}}        
\def\prc{\rm{Phys.~Rev.~C}}        
\def\prd{\rm{Phys.~Rev.~D}}        
\def\pre{\rm{Phys.~Rev.~E}}        
\def\prl{\rm{Phys.~Rev.~Lett.}}    
\def\pasp{\rm{PASP}}               
\def\pasj{\rm{PASJ}}               
\def\qjras{\rm{QJRAS}}             
\def\skytel{\rm{S\&T}}             
\def\solphys{\rm{Sol.~Phys.}}      
\def\sovast{\rm{Soviet~Ast.}}      
\def\ssr{\rm{Space~Sci.~Rev.}}     
\def\zap{\rm{ZAp}}                 
\def\nat{\rm{Nature}}              
\def\iaucirc{\rm{IAU~Circ.}}       
\def\aplett{\rm{Astrophys.~Lett.}} 
\def\apspr{\rm{Astrophys.~Space~Phys.~Res.}}
\def\bain{\rm{Bull.~Astron.~Inst.~Netherlands}} 
\def\fcp{\rm{Fund.~Cosmic~Phys.}}  
\def\gca{\rm{Geochim.~Cosmochim.~Acta}}   
\def\grl{\rm{Geophys.~Res.~Lett.}} 
\def\jcp{\rm{J.~Chem.~Phys.}}      
\def\jgr{\rm{J.~Geophys.~Res.}}    
\def\jqsrt{\rm{J.~Quant.~Spec.~Radiat.~Transf.}}
\def\memsai{\rm{Mem.~Soc.~Astron.~Italiana}}
\def\nphysa{\rm{Nucl.~Phys.~A}}   
\def\physrep{\rm{Phys.~Rep.}}   
\def\physscr{\rm{Phys.~Scr}}   
\def\planss{\rm{Planet.~Space~Sci.}}   
\def\procspie{\rm{Proc.~SPIE}}   

\let\astap=\aap
\let\apjlett=\apjl
\let\apjsupp=\apjs
\let\applopt=\ao

\maketitle

\begin{abstract}
We present the most detailed data-driven exploration of cloud opacity in a substellar object to-date. 
We have tested over 60 combinations of cloud composition and structure, particle size distribution, scattering model, and gas phase composition assumptions against archival $1-15 {\rm \mu m}$ spectroscopy for the unusually red L4.5~dwarf 2MASSW~J2224438-015852 using the {\it Brewster} retrieval framework.
We find that, within our framework, a model that includes enstatite and quartz cloud layers at shallow pressures, combined with a deep iron cloud deck fits the data best. This models assumes a Hansen distribution for particle sizes for each cloud, and Mie scattering. We retrieved particle effective radii of $\log_{10} a {\rm (\mu m)} = -1.41^{+0.18}_{-0.17}$ for enstatite, $-0.44^{+0.04}_{-0.20}$ for quartz, and $-0.77^{+0.05}_{-0.06}$ for iron. 
Our inferred cloud column densities suggest ${\rm (Mg/Si)} = 0.69^{+0.06}_{-0.08}$ if there are no other sinks for magnesium or silicon. 
Models that include forsterite alongside, or in place of, these cloud species are strongly rejected in favour of the above combination.
We estimate a radius of $0.75 \pm 0.02$~\Rjup, which is considerably smaller than predicted by evolutionary models for a field age object with the luminosity of 2M2224-0158.
Models which assume vertically constant gas fractions are consistently preferred over models that assume thermochemical equilibrium. 
From our retrieved gas fractions we infer ${\rm [M/H]} = +0.38^{+0.07}_{-0.06}$ and ${\rm C/O} = 0.83^{+0.06}_{-0.07}$. 
Both these values are towards the upper end of the stellar distribution in the Solar neighbourhood, and are mutually consistent in this context. 
A composition toward the extremes of the local distribution is consistent with this target being an outlier in the ultracool dwarf population.

\end{abstract}


\section{Introduction}
\label{sec:intro}

The importance of clouds for understanding the observed properties and evolution of giant (exo)planets and substellar objects is well established \citep[see e.g.][ for reviews]{sm08,helling2014,marley2015,helling2019,kirkpatrick2020}. 
This consensus view is underpinned by theoretical expectation and the features of substellar and exoplanet observations.
The first point boils down to recognising that condensation of various chemical species is thermodynamically favourable at pressures and temperatures found in a wide range of substellar and (exo)planetary atmospheres, although nucleation effects may also be important \citep{gao2020}. 
The second point draws on a varied set of direct and indirect inferences from data. 
At its most obvious, there is the observational fact of clouds seen across a wide range of atmospheres within the Solar System.

Beyond the Solar System, the inference of clouds has stemmed from their spectroscopic signatures.
In emission and transmission spectra alike, clouds mute the depth of features by providing continuum opacity that blocks flux through the otherwise brighter atmospheric windows between absorption bands.
The largest and most scientifically mature sample of objects displaying this kind of spectral signature of clouds are the L~dwarf population in the Solar neighbourhood, and it is these clouds that are the subject of this work. 
With temperatures in the range $1200 \ltsimeq T_{\rm eff} \ltsimeq 2300$K, their spectral sequence is widely understood as arising from the impact of cloud layers first appearing and then sinking beneath the photosphere with decreasing temperature \citep[e.g.][]{kirkpatrick2005}.

A range of different approaches for modelling clouds in substellar and exoplanet atmospheres are now described in the literature \citep[see e.g. ][for detailed reviews]{marley2015,helling2014}. 
These fall into two main categories: those that deal with the detailed microphysics of cloud nucleation and grain growth, and those that do not. 
The pioneering and popular \textsc{eddysed} cloud model \citep{ackerman2001} is an example of the latter.
This cloud model has been used widely in both self-consistent grid models and, more recently, in retrieval frameworks \citep[e.g.][]{sm08,morley2012,nowak2020,molliere2020}  
In this model, condensation of a particular species is assumed to take place at pressures and temperatures below condensation curves derived from phase equilibrium calculations and thermochemical model grids. 
These condensation curves are often plotted alongside thermal profiles in the literature to show potential cloud locations, including on the figures in this paper. 
The balance of mixing condensable gases upwards and cloud particles downwards is set by the sedimentation efficiency parameter ($f_{\rm sed}$), and grain sizes are set by mass balance considerations. A log-normal particle size distribution is assumed, with typical effective radii in the $\sim 1 - 10 \mu$m range arising. 
A similar approach is used by the BT~Settl model grid \citep[e.g. ][]{btsettlCS16}. However, in this case the condensation, sedimentation and mixing timescales are not parameterised but instead drawn from radiation–hydrodynamical simulations.

Other approaches are built around a detailed treatment of the microphysical processes that govern cloud particle nucleation and growth along with their sedimentation and diffusion \citep[e.g.][]{Helling2006,helling+2008,gao2020,powell2018}. 
These models predict a range of particle sizes, including significant fractions of sub-micron grains, and cloud compositions that diverge from predictions based on simple phase equilibrium.

In emission spectra, such as those of free-floating L~dwarfs, the effect of grey cloud can be mimicked by an isothermal temperature profile. 
This has led \citet{tremblin2015,tremblin2016} to suggest alternatives for the cloudy L~dwarf paradigm, wherein a chemical convective instability decreases the temperature gradient in the photosphere as the carbon chemistry moves from ${\rm CO}$ to ${\rm CH_4}$ dominated with decreasing temperature through the L~sequence to the T~dwarf sequence. 
However, while a log-normal size distribution with $\sim 1 - 10 \mu$m effective radius typically seen in the \citet{ackerman2001} cloud model produces a roughly grey opacity for most species, smaller particles and tighter size distributions produce non-grey opacity that can be distinguished from the cloud-free model, and allows different cloud species to be distinguished.  

In \citet[][hereafter B17]{burningham2017} we introduced the first retrieval analysis of cloudy L~dwarfs, and considered simple cloud parameterisations that allowed for grey and non-grey cloud opacity. 
The latter was treated as a simple function of wavelength raised to some power, i.e. $\tau \propto \lambda^{\alpha}$, where the power index was a retrieved parameter. 
Here, we build upon this work and extend our cloud model to incorporate specific cloud species via Mie scattering in an effort to make use of the full set of $1-15~\mu$m spectroscopy available for a range of L and T~dwarfs in archival data, including one of the targets from B17.
There are three main goals of this extension : 1) to resolve the significant disagreement that was found between the retrieved thermal profiles and those of the self-consistent model grids in B17; 2) to gain empirical insight into the cloud properties of an L~dwarf with respect to species and particle sizes; 3) to explore the impact of cloud treatment on estimates of composition indicators such as C/O ratio.

In this work we take a deep dive into the clouds of the red L~dwarf  2MASSW~J2224438-015852 (2M2224-0158 from here-on), exploring a wide-range of plausible (and implausible) cloud species and combinations thereof.
In Section~\ref{sec:target} we review the literature concerning 2M2224-0158. Section~\ref{sec:brew} outlines our retrieval framework, followed by a discussion of our model selection in Section~\ref{sec:res}. 
We provide analysis of the winning model and associated parameter estimates in Section~\ref{sec:prefmod}, and discuss the implications of our results in Section~\ref{sec:disc}. Conclusions are summarised in Section~\ref{sec:conc}.


\section{The target: 2M2224-0158}
\label{sec:target}
2M2224-0158 was originally identified in a search for L~dwarfs in the Two Micron All Sky Survey (\citet{kirkpatrick2000}).  Using a Keck LRIS spectrum the object was optically classified as an L4.5~dwarf with detectable H$\alpha$ emission.  Given the activity signature of the source, 2M2224-0158 was photometrically monitored and found to be variable at $I$ band \citep{Gelino_2002}, an early but strong signature of atmosphere dynamics.  It was also observed for linear polarization in search of a signature of dust in the atmosphere but no detection was made \citet{Menard2002,ZapateroOsorio05}.  \citet{dahn02} reported a parallax for the object making it an ideal candidate for numerous detailed studies of brown dwarfs.  
Importantly, \citet{cushing2005} examined the 0.6 - 4.1 $\mu$m spectrum and determined that it had an anomalously red spectrum typically interpreted as indicative of unusually thick condensate clouds and/or a low surface gravity. \citet{Cushing2009} followed up on that result, obtaining 5.5 - 38 $\mu$m data using the Spitzer Space Telescope's Infrared Spectrograph (IRS) and determined that there was a significant deviation from the model prediction at $\sim$ 9$\mu$m indicating the presence of a silicate feature.  
They speculated that this might be due to enstatite and/or forsterite clouds, while \citet{Helling2006} suggested that the feature could arise from quartz (${\rm SiO_2}$) grains, in contrast to  phase-equilibrium predictions \citep{visscher2010a}. 
\citet{sorahana2014} investigated the 2.5 - 5.0$\mu$m AKARI spectrum of 2M2224-0158 and found that their self-consistent grid models were unable to fit the $1 - 5~\mu$m spectrum without additional heating to raise the temperature in the upper ($P < 0.1~{\rm bar}$) atmosphere by a few 100K relative to equilibrium predictions.  
The consensus of literature conclusions on the atmosphere of 2M2224-0158 is that it is an outlier compared to the field population.

The anomalous red spectrum also begged the question as to whether or not the source was young.  Several groups investigated youth characteristics of this object in context with a growing sample of unusually red L dwarfs with spectroscopic signatures of a low surface gravity (e.g. 2M0355; \citealt{faherty2013} PSO 318; \citealt{liu2013}, etc.).  For instance, \citet{Gagne14}, \citet{Faherty_2016}, \citet{Martin_2017} and \citet{Liu_2016} examined the kinematics, near infrared spectrum, and/or color magnitude position of 2M2224-0158 and concluded that it was a field gravity L~dwarf with space motion consistent with the old field population near the Sun.  At present there is no obvious moving group or association that 2M2224-0158 might belong to.  Therefore the anomalous observables are speculated to be atmospheric in nature and not a consequence of youth.

B17 performed the first atmospheric retrieval of 2M2224-0158.  That work obtained effective temperature and $\log g$ results which matched semi-empirical results using just the near infrared spectrum of the source.  Importantly for this work, B17 applied 6 different approaches to the cloud formulation of 2M2224-0158 and found that a power law cloud deck was the best fitting model. 
The optically thick cloud deck which passes $\tau_{cloud} \geq 1$ (looking down) at a pressure of around 5~bar. The temperature at this pressure is too high for silicate species to condense, and B17 argued
that corundum and/or iron clouds are responsible for this cloud opacity. The retrieved profiles were cooler at depth and warmer at altitude than the forward grid model comparisons, therefore B17 speculate that some form of heating mechanism may be at work in the upper atmospheres of 2M2224-0158. All of these conclusions were drawn in the absence of the mid-infrared spectrum which this paper now includes in the analysis.

\begin{table*}
\begin{tabular}{c c c c}
\hline
Parameter & Value & Notes / units & Reference \\
\hline
$\alpha$ &  22:24:43.8 & J2000 & \citet{gaiadr2}\\
$\delta$ &  -01:58:52.14 & J2000 & \citet{gaiadr2}\\
 Distance & $11.56 \pm 0.10$ & parsec & \citet{gaiadr2}\\
$\mu_{\alpha \cos \delta}$ & $471.049 \pm 0.770$ & mas/yr &  \citet{gaiadr2} \\
$\mu_\delta$ & $-874.914 \pm 0.821$ & mas/yr &  \citet{gaiadr2} \\
$v_{rad}$ & $-36.48 \pm 0.01$ & ${\rm km s^{-1}}$ & \citet{Faherty_2016} \\
Spectral type & L3.5 & optical & \citet{stephens2009} \\
& L4.5 & near-infrared & \citet{stephens2009} \\
Spectra & SpeX & $1 - 2.5~\mu$m & \citet{burgasser2010} \\
& IRCS $L'$ band &  $2.5 - 5~\mu$m & \citet{cushing2005} \\
& {\it Spitzer} IRS & $5 - 15~\mu$m &  \citet{cushing2006} \\
$G$ &  $19.3686 \pm 0.0053$ & mag & \citet{gaiadr2} \\
$J$ & $14.073 \pm 0.027$ & mag & \citet{2mass} \\
$H$ & $12.818 \pm 0.026$ &  mag & \citet{2mass} \\
$K_s$ & $12.022 \pm 0.023$ & mag & \citet{2mass} \\
$\log (L_{\rm bol} / \Lsun)$ &  $-4.16 \pm 0.01$ & empirical & \citet{filippazzo2015} \\
 &  $-4.146 \pm 0.003$ & inferred & this work \\
Mass & $60.57 \pm 15.26$ & ${\rm M_{Jup}}$ (semi-empirical) & \citet{filippazzo2015} \\
& $67^{+11}_{-14}$ & ${\rm M_{Jup}}$ / inferred & this work \\
Radius & $0.99 \pm 0.08$ & ${\rm R_{Jup}}$ (semi-empirical) & \citet{filippazzo2015} \\
& $0.75 \pm 0.02$ & ${\rm R_{Jup}}$ (inferred) & this work \\
$T_{\rm eff}$ & $1646 \pm 71$ & K (semi-empirical) & \citet{filippazzo2015} \\
& $1912^{+18}_{-19}$ & K (inferred) & this work \\
$\log g$ & $5.18 \pm 0.22$ & semi-empirical & \citet{filippazzo2015} \\
& $5.47^{+0.07}_{-0.11}$ & retrieved & this work \\
${\rm [M/H]}$ & $0.38^{+0.07}_{-0.06}$ & inferred &  this work \\
C/O & $0.83^{+0.06}_{-0.07}$ & inferred & this work \\
\hline
\end{tabular}
\caption{Summary of the available parameters and data used for our analysis of 2M2224-0158. Also included are bulk properties that have been retrieved or inferred from our retrieved parameters. 
\label{properties}
}
\end{table*}

\section{Retrieval framework}
\label{sec:brew}
Our retrieval framework (nicknamed ``{\it Brewster}'') is an extension of that described in B17, which we have improved in respect to our cloud model and our treatment of gas phase abundances and opacities. For a more complete description of the framework and its validation we refer the reader to B17. 
We summarise its key features before discussing the new extensions in more detail below.

Our radiative transfer scheme evaluates the emergent flux from a layered atmosphere in the two stream source function technique of \citet{toon1989}, including scattering, as first introduced by \citet{mckay1989} and subsequently used by e.g. \citet{marley1996,sm08,morley2012}. 
In {\it Brewster's} default arrangement (used here), we set up a 64 layer atmosphere (65 levels) with geometric mean pressures in the range $\log P({\rm bar}) = -4$ to 2.3, spaced at 0.1~dex intervals.  

We set the temperature in each layer via the multiple exponential parameterisation put forward by \citet{madhu2009}. This scheme treats the atmosphere as three zones: 

\begin{equation}
\begin{aligned}
P_{0} < P < P_{1}: P  = P_{0} e^{\alpha_{1}(T - T_{0})^{\frac{1}{2}}}   \hfill (\text{Zone 1})\\
P_{1} < P < P_{3}: P  = P_{2} e^{\alpha_{2}(T - T_{2})^{\frac{1}{2}}}   \hfill (\text{Zone 2})\\
P > P_{3} : T = T_{3}  \hfill (\text{Zone 3})
\end{aligned}
\label{eqn:madhu}
\end{equation}
where $P_{0}$, $T_{0}$ are the pressure and temperature at the top of the atmosphere, which becomes isothermal with temperature $T_{3}$ at pressure $P_{3}$.
In its most general form, a thermal inversion occurs when $P_{2} > P_{1}$. Since $P_{0}$ is fixed by our atmospheric model, and continuity at the zonal boundaries allows us to fix two parameters, we consider six free parameters $\alpha_{1}$, $\alpha_{2}$, $P_1$, $P_2$, $P_3$, and $T_3$.  If we rule out a thermal inversion by setting $P_{2} = P_{1}$ \citep[see Figure 1, ][]{madhu2009},  we can further simplify this to five parameters $\alpha_{1}$, $\alpha_{2}$, $P_1$, $P_3$, $T_3$.

In this work we consider the following absorbing gases: ${\rm H_2O, CO, CO_2, CH_4, TiO, VO, CrH, FeH, Na, K}$.
These gases were chosen as they have been previously identified as important absorbing species in mid-L~dwarf spectra, and which thus are amenable to retrieval analysis.

We calculate layer optical depths due to these absorbing gases using opacities sampled at a resolving power R = 10,000, which are taken from the compendium of \citet{freedman2008,freedman2014}, with updated opacities described in B17. 
The line opacities are tabulated across our temperature-pressure regime in 0.5~dex steps for pressure, and with temperature steps ranging from 20~K to 500~K as we move from 75~K to 4000~K. We then linearly interpolate to our working pressure grid.  

We include continuum opacities for H$_{2}$-H$_{2}$ and H$_{2}$-He collisionally induced absorption, using the cross-sections from  \citet{richard2012} and \citet{saumon2012}, and Rayleigh scattering due to H$_{2}$, He and CH$_{4}$ but neglect the remaining gases.
We also include continuum opacities due to bound-free and free-free absorption by H$^-$ \citep{john1988, Bell+1987}, and free-free absorption by H$_2^-$ \citep{bell1980}.

The D resonance doublets of NaI ($\sim 0.59~\mu$m) and KI ($\sim 0.77~\mu$m) can become extremely strong in the spectra of brown dwarfs, with line profiles detectable as far as $\sim$3000~cm$^{-1}$ from the line centre in T dwarfs \citep[e.g. ][]{burrows2000,liebert2000,marley2002,king2010}. Under these circumstances, a Lorentzian line profile becomes woefully inadequate in the line wings and a detailed calculation is required. For these two doublets, we have implemented line wing profiles based on the unified line shape theory \citep{nallard2007a,nallard2007b}. 
The tabulated profiles (Allard N., private communication) are calculated for the D1 and D2 lines of \ion{Na}{I} and \ion{K}{I} broadened by collisions with H$_{2}$ and He, for temperatures in the $500 - 3000$~K range and perturber (H$_{2}$ or He) densities up to $10^{20}$ cm$^{-3}$. Two collisional geometries are considered for broadening by H$_{2}$. The profile within 20~cm$^{-1}$ of the line centre is Lorentzian with a width calculated from the same theory.


We estimate parameters and compare models in a Bayesian framework, and sample the posterior probabilities using the \textsc{emcee} \citep{emcee} algorithm. 
As with B17, we use 16 walkers per dimension, and extend our iterations until we have run for at least fifty times the estimated autocorrelation length. 
We also check that the maximum likelihood is no longer evolving with additional sets of 10,000 -- 30,000 iterations. 
Typically our \textsc{emcee} chains run for 70,000 -- 130,000 iterations.  
We initialise all of our model cases identically, using tight Gaussians centered on  equilibrium predictions for solar composition gas abundances and $\log g = 5.0$. 
Our clouds are initialised with broader distributions reflecting our ignorance of cloud locations, but roughly corresponding to the locations of condensation curves along a $T_{\rm eff} = 1700$~K $\log g = 5.0$ \citet{sm08} grid model thermal profile. 
The thermal parameters are also loosely initialised around the same profile. 
Since we are combining data from multiple instruments, we also allow for calibration errors by including relative scale factors between the SpeX data and the GNIRS and {\it Spitzer} data, respectively.  
Data from each instrument also has its own tolerance parameter to reflect the range of SNR exhibited by these data sets. 
We use uniform and log-uniform priors which generously enclose the plausible parameter space. These are summarised in Table~\ref{tab:priors}. 
The full set of free-parameters in a particular model proposal is called the state vector.

\begin{table*}
\begin{tabular}{cc}
\hline
Parameter & Prior \\
\hline
gas fraction ($X_{gas}$) & log-uniform, $\log X_{gas} \geq -12.0$, $\sum_{gas}{X_{gas}} \leq 1.0$ \\
thermal profile:  $\alpha_{1}, \alpha{2}, P1, P3, T3$ & uniform, constrained by $0.0~{\rm K} < T_{i} < 5000.0~{\rm K}$ \\
scale factor, $R^{2} / D^{2}$ & uniform, constrained by $0.5 R_{Jup} \leq R \leq 2.0 R_{Jup}$ \\
gravity, $\log g$ & uniform, constrained by $1M_{Jup}  \leq gR^{2} / G \leq 80M_{Jup}$\\ 
cloud top$^{1}$, $P_{top}$ & log-uniform, $-4 \leq \log_{10} P_{top}  \leq +2.3$ \\
cloud decay scale$^{2}$, $(\Delta \log_{10} P)_{decay}$  & log-uniform, $0 < (\Delta \log_{10}  P)_{decay} < 7$ \\
cloud thickness$^{3}$ $(\Delta \log_{10} P)_{thick}$ & log-uniform, constrained by $\log_{10} P_{top} \leq \log_{10} P_{top} + (\Delta \log_{10} P)_{thick} \leq 2.3$\\
cloud total optical depth (extinction) at 1 $\mu$m$^{3}$ & uniform, $0.0 \leq \tau_{cloud} \leq 100.0$ \\
Hansen distribution effective radius, $a$ & log-uniform, $-3.0 < \log_{10} a {\rm (\mu m)} < 3.0$ \\
Hansen distribution spread, $b$ & uniform, $0 < b < 1.0$ \\
Geometric mean radius (log-normal distribution), $\mu$ & $-3.0 < \log_{10} \mu {\rm (\mu m)} < 3.0$ \\
Geometric standard deviation (log-normal distribution), $\sigma$ &  $1 < \sigma < 5$ \\
wavelength shift, $\Delta lambda$ & uniform, $-0.01< \Delta \lambda < 0.01 \micron$ \\
tolerance factor, $b$ & uniform, $\log (0.01 \times min(\sigma_{i}^{2})) \leq b \leq \log(100 \times max(\sigma_{i}^{2}))$ \\
\hline
\end{tabular}
\caption{Notes: 1) for an optically thick cloud deck this is the pressure where $\tau_{cloud} = 1$, for a slab cloud this is the top level of the slab; 2) decay height for cloud deck above the $\tau_{cloud} = 1.0$ level; 3) thickness and $\tau_{cloud}$ only retrieved for slab cloud.   \label{tab:priors} }
\end{table*}

\subsection{Gas phase abundances}
\begin{figure}
\begin{center}
\hspace{0cm}
\includegraphics[width=225pt]{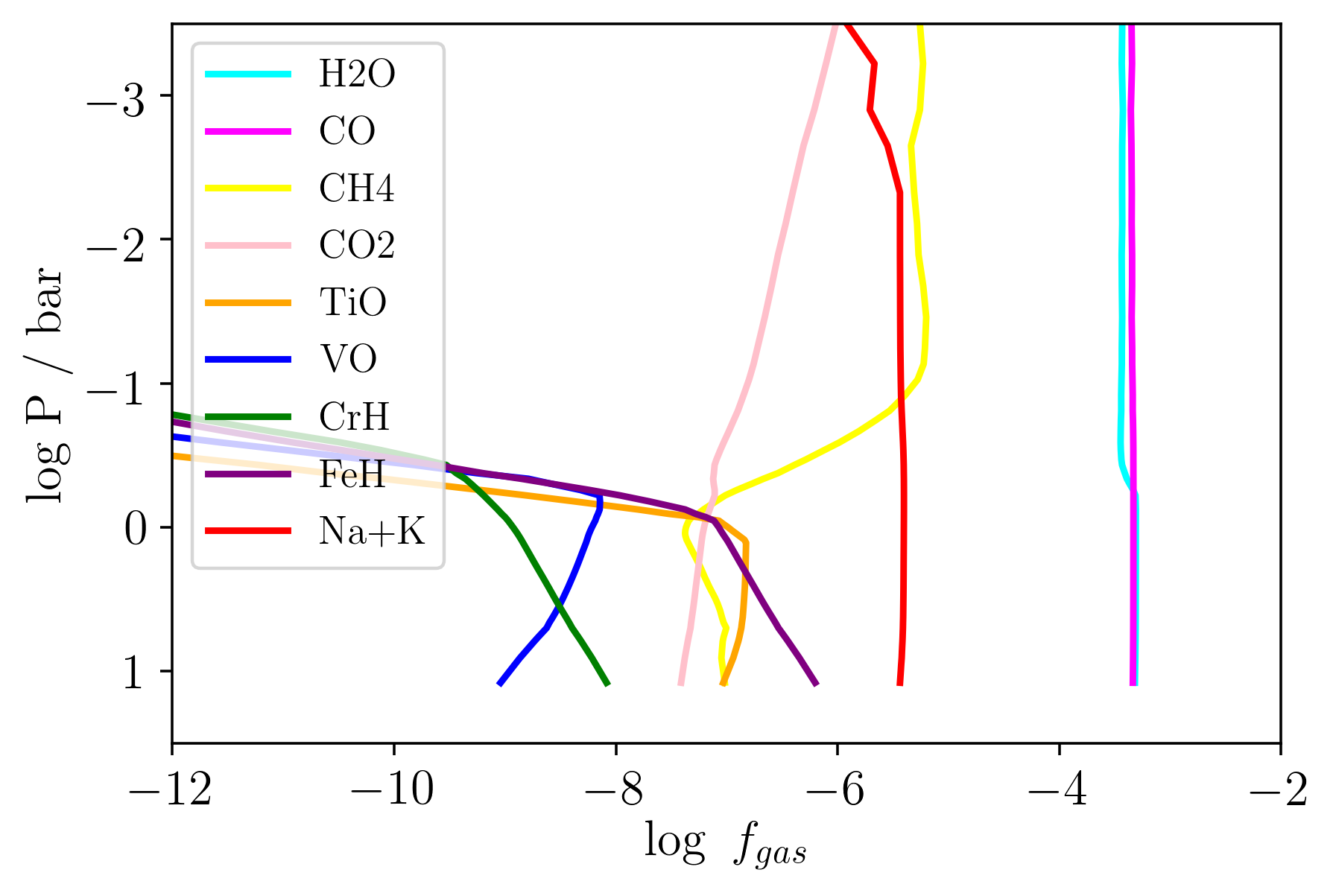}
\caption{Predicted gas fractions ($f_{\rm gas}$) from our thermochemical grid calculated along a grid model \teff~=~1700~K $\log g = 5.0$ thermal profile. 
\label{fig:chemeq}}
\end{center}
\end{figure}

In this work we incorporate two methods for handling gas phase abundances. 
The first method is to specify vertically constant volume mixing fractions for each gas, and retrieve these directly, as was done in B17. This is a common approach in so-called ``free" retrievals. 
This method has the advantage of simplicity, but may fail to capture important variations of gas abundances with altitude.
Figure~\ref{fig:chemeq} shows expected thermochemical equilibrium abundances for likely absorbing gases in the LT regime, calculated along a thermal profile taken from a $T_{\rm eff} = 1700$K, $\log g = 5.0$, solar metallicity self-consistent Marley \& Saumon grid model. It can be seen that abundances for many of our absorbers are expected to vary by several orders of magnitude within the 0.1 - 10 bar region from which we expect to have large contributions of flux. 
Although free retrieval of gas abundances that vary with altitude may be desirable, the large number of parameters involved (e.g. 5+ per gas to provide meaningful flexibility), the necessity of allowing sharp changes in vertical profile, and the potential degeneracy with the thermal profile suggest such an approach to the inverse problem would be ill-posed.

The second method we use here attempts to address this shortcoming by assuming thermochemical equilibrium, and retrieving [Fe/H] and C/O instead of individual gas abundances. The gas fractions in each layer are then drawn from tables of thermochemical equilibrium abundances as a function of T, P, [Fe/H], C/O ratio along the thermal profile of a given state vector.

The thermochemical equilibrium grids were calculated using the NASA Gibbs minimization CEA code \citep[see  ][]{mcbride1994}, based upon prior thermochemical models \citep{fegley1994,fegley1996,lodders1999,lodders2002,lodders2010,lodders2002b,lodders2006,visscher2006,visscher2010a,visscher2012,moses2012,moses2013} and recently utilized to explore gas and condensate chemistry over a range of substellar atmospheric conditions \citep{morley2012,morley2013,skemer2016,kataria2016,wakeford2016}.
The chemical grids are used to determine the equilibrium abundances of atmospheric species over a wide range of atmospheric pressures (from 1 microbar to 300 bar), temperatures (300 to 4000 K), metallicities ($-1.0 < {\rm [Fe/H]} < +2.0$),  and C/O abundance ratios (0.25 to 2.5 times the solar C/O abundance ratio).

\subsection{Cloud model}
\label{sec:clouds}
Our cloud model requires two levels of specification: 1) the cloud structure and location in the atmosphere; 2) optical properties of the cloud particles that define the wavelength dependence of its opacity. 
For the former, we adopt the same methods for parameterising cloud structure as we did in B17. 
Each cloud is approximated as one of two options: a ``slab'' cloud or a ``deck'' cloud.
Both are defined by the manner in which opacity due to the cloud is distributed amongst the layers of the atmosphere.
The slab cloud has the following parameters: a cloud top pressure ($P_{top}$), a physical extent in log-pressure ($\Delta \log P$), and a total optical depth at 1$\mu$m ($\tau_{cloud}$). 
The optical depth of the slab cloud is distributed through its extent as $d\tau / dP \propto P$ (looking down), reaching its total value at the high-pressure extent of the slab. This cloud can have any optical depth in principle, but we restrict the prior to $0.0 \leq \tau_{cloud} \leq 100.0$.

The deck cloud differs from the slab cloud in that it {\it always becomes optically thick at some pressure}. If the slab cloud can be thought of as a cloud that we {\it might} be able to see to the bottom of, the deck cloud is one that we only see the top of, and thus cannot infer the true total optical depth. Instead, it is parameterised by the pressure at which its total optical depth at 1$\mu$m passes unity (looking down) for the first time ($P_{deck}$), and a decay height $\Delta \log P$ over which the optical depth falls to shallower pressures as $d\tau/dP \propto \exp((P-P_{deck}) / \Phi)$, where\\

\begin{equation}
\begin{aligned}
\Phi = \frac{P_{top}(10^{\Delta \log P} - 1)}{10^{\Delta \log P}}
\end{aligned}
\label{eqn:deck}
\end{equation}.

The optical depth of the deck cloud increases following the same function to deeper pressures until $\Delta \tau_{layer} = 100$.  Deck clouds can become opaque very rapidly with increasing pressure, such that essentially no information about the atmosphere from deep beneath the deck is accessible. This must be borne in mind when interpreting retrieved (parameterised) thermal profiles. In deck cloud cases, the profile below the deck (and its spread) simply extends the gradient of the profile at the cloud deck (and spread therein).

We consider the wavelength dependent optical properties of different condensate species under the assumption of Mie scattering, and also test the alternative model of Distributed Hollow Spheres \citep{min2005} for a small subset of models.  
We have taken the optical data (refractive indices) for likely condensates from a variety of sources (see Table~\ref{tab:optics}), and have pre-tabulated Mie coefficients for our condensates as a function of particle radius and wavelength.

\begin{table}
\begin{center}
\begin{tabular}{c c}
\hline
Condensate & Reference \\
\hline
${\rm Al_2 O_3}$ & \citet{al2o3} \\
Fe & \citet{leksina1967} \\
amorphous ${\rm Mg SiO_3}$ &  \citet{scott1996} \\
crystalline ${\rm Mg SiO_3}$ & \citet{jaeger1998} \\
amorphous ${\rm Mg_2 SiO_4}$ & \citet{scott1996} \\
crystalline ${\rm Mg_2 SiO_4}$ & \citet{servoin1973} \\
Fe-rich ${\rm Mg_2 SiO_4}$ & \citet{wakeford2015}\\
${\rm Fe_2 O_3}$ & \citet{wakeford2015} \\
${\rm Si O_2}$ & \citet{wakeford2015} \\
${\rm Mg Al_2 O_4}$ & \citet{wakeford2015} \\
${\rm Ti O_2}$ & \citet{wakeford2015} \\
${\rm Ca Ti O_3}$ & \citet{wakeford2015} \\
\hline
\end{tabular}
\caption{Sources for optical data for condensates used in this work
\label{tab:optics}
}
\end{center}
\end{table}

Wavelength dependent optical depths, single scattering albedos and phase angles in each layer are then calculated at runtime by integrating the cross sections and Mie efficiencies over the particle size distribution in that layer for each cloud species present. 
The total particle number density for a given condensate in a layer is calibrated to the optical depth at 1$\mu$m in the layer as determined by the parameterised cloud structure in the state vector. 
In this work we explore retrieval models that assume either a \citet{hansen1971} or a log-normal distribution of particle sizes. \\
For a Hansen distribution the number $n$ of particles with radius $r$ is defined as:

\begin{equation}
\begin{aligned}
n(r) \propto r^\frac{1-3b}{b}e^{-\frac{r}{ab}}
\end{aligned}
\label{eqn:hansen}
\end{equation}

where the parameters $a$ and $b$ refer to the effective radius and spread of the distribution respectively, and are given by:\\

\begin{equation}
\begin{aligned}
a = \frac{\int_{0}^{\infty}r \pi r^2 n(r)dr}{\int_{0}^{\infty}\pi r^2 n(r) dr}
\end{aligned}
\label{eqn:hana}
\end{equation}

\begin{equation}
\begin{aligned}
b = \frac{\int_{0}^{\infty} (r - a)^2 \pi r^2 n(r) dr}{a^2 \int_{0}^{\infty} \pi r^2 n(r) dr}
\end{aligned}
\label{eqn:hanb}
\end{equation}
.

In {\it Brewster}, we are able to simulate opacity arising by combinations of condensate species by combining one of our simple cloud structures with a set of optical properties to define a given ``cloud''. We can then incorporate any linear combination of these ``clouds" into our retrieval model to arbitrary degrees of complexity.

\section{Model Selection}
\label{sec:res}

In an effort to explore as much potential cloud parameter space as possible in limited computation time, we have simulated a range of structural combinations, building up in complexity from a simple no-cloud model to combinations of up to four cloud species. These arrangements typically combine one or more slabs with a single deck cloud. We have not modelled multiple species of deck cloud. This is because any cloud opacity that lies at deeper pressure than a deck cloud's optically thick region will be totally obscured. 
Since a slab cloud is capable of reproducing much of the behaviour of a deck cloud, we only employ a single deck cloud in each model, which is typically retrieved as the deepest cloud of the set (although it is free to appear anywhere with atmosphere according to the prior).   
Each cloud structure represents a single cloud species, and different slabs and a deck are free to co-locate in pressure. 
Our choices of model fall into two groups based on the rationale for selecting the model. 
These are:\\
\begin{enumerate}
    \item expected species based on phase equilibrium and cloud modelling predictions
    
    \item species selected on basis of Mie-scattering features that overlap with features seen in the spectrum of 2M2224-0158.
    
\end{enumerate}

It is worth noting that the second group may include implausible species. That is, species that we don't expect given our understanding of substellar atmospheres e.g. ${\rm Fe_2 O_3}$, which is very unlikely to be found in such a reducing environment.

In B17 we found that the cloud opacity seen there was most consistent with a Hansen distribution of particle sizes (see Equations~\ref{eqn:hansen},~\ref{eqn:hana},~\ref{eqn:hanb}). 
For this reason, we have focused our computational resources on models that assumes this particle size distribution, and treat scattering with Mie theory. 
However, for completeness we have also tested a subset of models using assuming a log-normal particle size distribution, and also DHS scattering (with Hansen distribution).
In addition, for a subset of cloud arrangements focused on the top-ranked models at different levels of complexity, we tested models that assume thermochemical equilibrium gas abundances.

Table~\ref{tab:bics} lists the wide range of combinations of condensate species and arrangements that we have modelled in our retrievals. 
We rank our models according to the Bayesian Information Criterion (BIC), in order of increasing $\Delta BIC$ from the ``winning" model. 
The BIC provides a method for model selection for cases when it is not possible to calculate the Bayesian Evidence for a set of models. The BIC is defined as:

\begin{equation}
\begin{aligned}
BIC = k \ln(n) - 2 \ln(\likL)
\end{aligned}
\label{eqn:bic}
\end{equation}

where: $k =$~number of parameters; $n =$~the number of data points; $\likL =$~the likelihood. \citet{kass1995} provide the following intervals for selecting between two models under the BIC:

\begin{itemize}
  \item[] $0  < \Delta BIC < 2$: no preference worth mentioning;
  \item[] $2  < \Delta BIC < 6$: postive;
  \item[] $6  < \Delta BIC < 10$: strong;
 \item[] $10  < \Delta BIC$: very strong.
\end{itemize}

\begin{table*}
\begin{tabular}{c c c c c c}
\hline
Cloud 1 & Cloud 2  & Cloud 3 & Cloud 4 & $N_{parameters}$ (notes) & $\Delta BIC$ \\
\hline
am-${\rm MgSiO_3}$ slab &  ${\rm SiO_2}$ slab & Fe deck & N/A & 36 & 0 \\
am-${\rm MgSiO_3}$ slab & ${\rm Fe_2 O_3}$ slab & Fe deck & N/A & 36 & 15 \\
am-${\rm MgSiO_3}$ slab & am-${\rm Mg_2 SiO_4}$ slab & Fe deck & N/A & 36 & 21 \\
am-${\rm MgSiO_3}$ slab &  ${\rm SiO_2}$ slab &  am-${\rm Mg_2 SiO_4}$ deck & N/A & 36 & 22 \\
am-${\rm MgSiO_3}$ slab &  ${\rm SiO_2}$ slab & ${\rm Al_2 O_3}$ deck & N/A & 36 & 23 \\
am-${\rm MgSiO_3}$ slab & cry-${\rm MgSiO_3}$ slab & Fe deck & N/A & 36 & 46 \\
am-${\rm MgSiO_3}$ slab & ${\rm SiO_2}$ slab & am-${\rm Mg_2 SiO_4}$ slab & Fe deck & 41 & 57 \\
cry-${\rm MgSiO_3}$ slab & ${\rm SiO_2}$ slab & ${\rm TiO_2}$ deck & N/A & 36	& 88 \\
am-${\rm MgSiO_3}$ & ${\rm Fe_2 O_3}$ deck & N/A & N/A & 31 & 96 \\
am-${\rm MgSiO_3}$ slab & Fe deck & N/A & N/A & 31 & 101 \\
cry-${\rm MgSiO_3}$ slab & ${\rm SiO_2}$ slab & ${\rm Fe_2 O_3}$ slab & Fe deck & 41 & 104 \\
cry-${\rm MgSiO_3}$ slab & ${\rm SiO_2}$ slab & cry-${\rm Mg_2 SiO_4}$ slab & Fe deck & 41 & 106 \\
cry-${\rm MgSiO_3}$ slab & ${\rm SiO_2}$ slab & ${\rm Fe_2 O_3}$ deck & N/A & 36 & 115 \\
cry-${\rm MgSiO_3}$ slab & ${\rm SiO_2}$ slab & ${\rm Mg Al_2 O_4}$ deck & N/A & 36 & 122 \\
cry-${\rm MgSiO_3}$ slab & ${\rm Fe_2 O_3}$ slab & Fe deck & N/A & 36 & 122 \\
cry-${\rm MgSiO_3}$ slab &  ${\rm SiO_2}$ slab & Fe deck  & N/A & 36 & 129 \\
am-${\rm MgSiO_3}$ slab & ${\rm Al_2 O_3}$ deck & N/A & N/A & 31 & 140 \\
cry-${\rm MgSiO_3}$ slab & ${\rm SiO_2}$ slab & ${\rm Fe_2 O_3}$ slab &  ${\rm Mg Al_2 O_4}$ deck & 41 & 179 \\
cry-${\rm MgSiO_3}$ slab &  ${\rm SiO_2}$ slab & ${\rm Al_2 O_3}$ deck & N/A & 36 & 179 \\
cry-${\rm MgSiO_3}$ slab & ${\rm Fe_2 O_3}$ deck & N/A & N/A & 31 & 195 \\
am-${\rm MgSiO_3}$ deck & N/A & N/A & N/A & 26	& 207 \\
cry-${\rm MgSiO_3}$ slab &  ${\rm SiO_2}$ deck & N/A & N/A & 31 & 216 \\
Fe-rich ${\rm Mg_2 SiO_4}$ slab & Fe deck & N/A & N/A & 31 & 217 \\
cry-${\rm MgSiO_3}$ slab & Fe deck & N/A & N/A & 31 & 218 \\
am-${\rm MgSiO_3}$ slab & N/A & N/A & N/A & 27 & 226 \\
cry-${\rm MgSiO_3}$ slab & ${\rm Mg Al_2 O_4}$ deck & N/A & N/A & 31 & 230 \\
cry-${\rm MgSiO_3}$ slab & ${\rm Al_2 O_3}$ deck & N/A & N/A & 31 & 233 \\
am-${\rm MgSiO_3}$ slab & Fe deck & N/A & N/A & 24 (CE) & 242 \\
am-${\rm MgSiO_3}$ slab & ${\rm Fe_2 O_3}$ slab & Fe deck & N/A & 29 (CE) & 263 \\
am-${\rm MgSiO_3}$ slab &  ${\rm SiO_2}$ slab & ${\rm Al_2 O_3}$ deck & N/A & 29 (CE) & 269 \\
am-${\rm MgSiO_3}$ slab & am-${\rm Mg_2 SiO_4}$ slab & Fe deck & N/A & 29 (CE) & 277 \\
am-${\rm MgSiO_3}$ slab &  ${\rm SiO_2}$ slab & Fe deck & N/A & 29 (CE) & 278 \\
am-${\rm MgSiO_3}$ slab &  ${\rm SiO_2}$ slab &  am-${\rm Mg_2 SiO_4}$ deck & N/A & 29 (CE) & 282 \\
cry-${\rm MgSiO_3}$ slab & ${\rm SiO_2}$ slab & ${\rm Mg Al_2 O_4}$ slab & Fe deck & 41 & 290 \\
am-${\rm MgSiO_3}$ slab & ${\rm Fe_2 O_3}$ deck  & N/A & N/A & 24 (CE) & 293 \\
am-${\rm MgSiO_3}$ deck & N/A & N/A & N/A & 19 (CE) & 319 \\
cry-${\rm MgSiO_3}$ slab &  ${\rm SiO_2}$ slab & ${\rm CaTiO_3}$ deck & N/A & 36 & 336 \\
am-${\rm MgSiO_3}$ slab & N/A & N/A & N/A & 20 (CE) & 377 \\
cry-${\rm MgSiO_3}$ slab &  cry-${\rm Mg_2 SiO_4}$ slab & Fe deck & N/A & 36 & 387 \\
am-${\rm MgSiO_3}$ slab &  ${\rm SiO_2}$ slab & ${\rm CaTiO_3}$ deck & N/A & 36 & 414 \\
am-${\rm MgSiO_3}$ slab &  ${\rm SiO_2}$ slab & am-${\rm Mg_2 SiO_4}$ deck & N/A & 36 (log-normal) & 440 \\
cry-${\rm MgSiO_3}$ slab & ${\rm Fe_2 O_3}$ deck & N/A & N/A & 24 (CE) & 527 \\
cry-${\rm MgSiO_3}$ slab &  cry-${\rm Mg_2 SiO_4}$ slab &  N/A & N/A & 32 & 531 \\
cry-${\rm MgSiO_3}$ slab & N/A & N/A & N/A & 27 & 545 \\
cry-${\rm MgSiO_3}$ deck & N/A & N/A & N/A & 26 & 547 \\
am-${\rm MgSiO_3}$ slab & ${\rm Fe_2 O_3}$ slab & Fe deck & N/A & 36 (DHS) & 548 \\
am-${\rm MgSiO_3}$ slab &  ${\rm SiO_2}$ slab & Fe deck & N/A & 36 (DHS) & 557 \\
cry-${\rm MgSiO_3}$ slab & ${\rm Fe_2 O_3}$ slab & N/A & N/A & 32 & 574 \\
cry-${\rm MgSiO_3}$ slab & Fe deck &  N/A & N/A & 31 (DHS) & 577 \\
am-${\rm MgSiO_3}$ slab & am-${\rm Mg_2 SiO_4}$ slab & Fe deck & N/A & 36 (DHS) & 604 \\
cry-${\rm MgSiO_3}$ slab & ${\rm Fe_2 O_3}$ deck &  N/A & N/A & 31 (DHS) & 608 \\
${\rm Fe_2 O_3}$ deck & N/A & N/A & N/A & 26 & 612 \\
am-${\rm MgSiO_3}$ slab &  ${\rm SiO_2}$ slab &  am-${\rm Mg_2 SiO_4}$ deck & N/A & 36 (DHS) & 695 \\
No cloud & N/A & N/A & N/A & 22 & 709 \\
${\rm Fe_2 O_3}$ slab & N/A & N/A & N/A & 27 & 735 \\
No cloud & N/A & N/A & N/A & 15 (CE) & 908 \\
am-${\rm MgSiO_3}$ slab &  ${\rm SiO_2}$ slab & ${\rm Al_2 O_3}$ deck & N/A & 36 (log-normal) & 1245 \\
am-${\rm MgSiO_3}$ slab & Fe deck & N/A & N/A & 31 (log-normal) & 1491 \\
am-${\rm MgSiO_3}$ slab &  ${\rm SiO_2}$ slab & Fe deck & N/A & 36 (log-normal) & 1752 \\
am-${\rm MgSiO_3}$ slab & am-${\rm Mg_2 SiO_4}$ slab & Fe deck & N/A & 36 (log-normal) & 1843 \\
am-${\rm MgSiO_3}$ slab & ${\rm Fe_2 O_3}$ slab & Fe deck & N/A & 36 (log-normal) & 1874 \\

\hline
\hline
\end{tabular}
\caption{List of models tested in this work for 2M2224-0158 along with $\Delta BIC$. Rows with no additional notes alongside the number of parameters are model runs with our default vertically constant gas fractions, Hansen particle size distribution, and Mie scattering. Changes to this are indicated in brackets as follows: CE = chemical equilibrium gas fractions; log-normal = log-normal particle size distribution; DHS = distribution of hollow spheres scattering.  
\label{tab:bics}}
\end{table*}

\section{Results \& Analysis}
\label{sec:prefmod}
The top ranked model is one that combines slabs of enstatite (${\rm MgSiO_3}$) and quartz (${\rm SiO_2}$), with a deeper iron deck.  The $\Delta BIC$ suggests that this model is strongly preferred over other models, with the closest ``runner up" having a  $\Delta BIC = 15$.

\begin{figure*}
\hspace{-0.8cm}
\includegraphics[width=300pt]{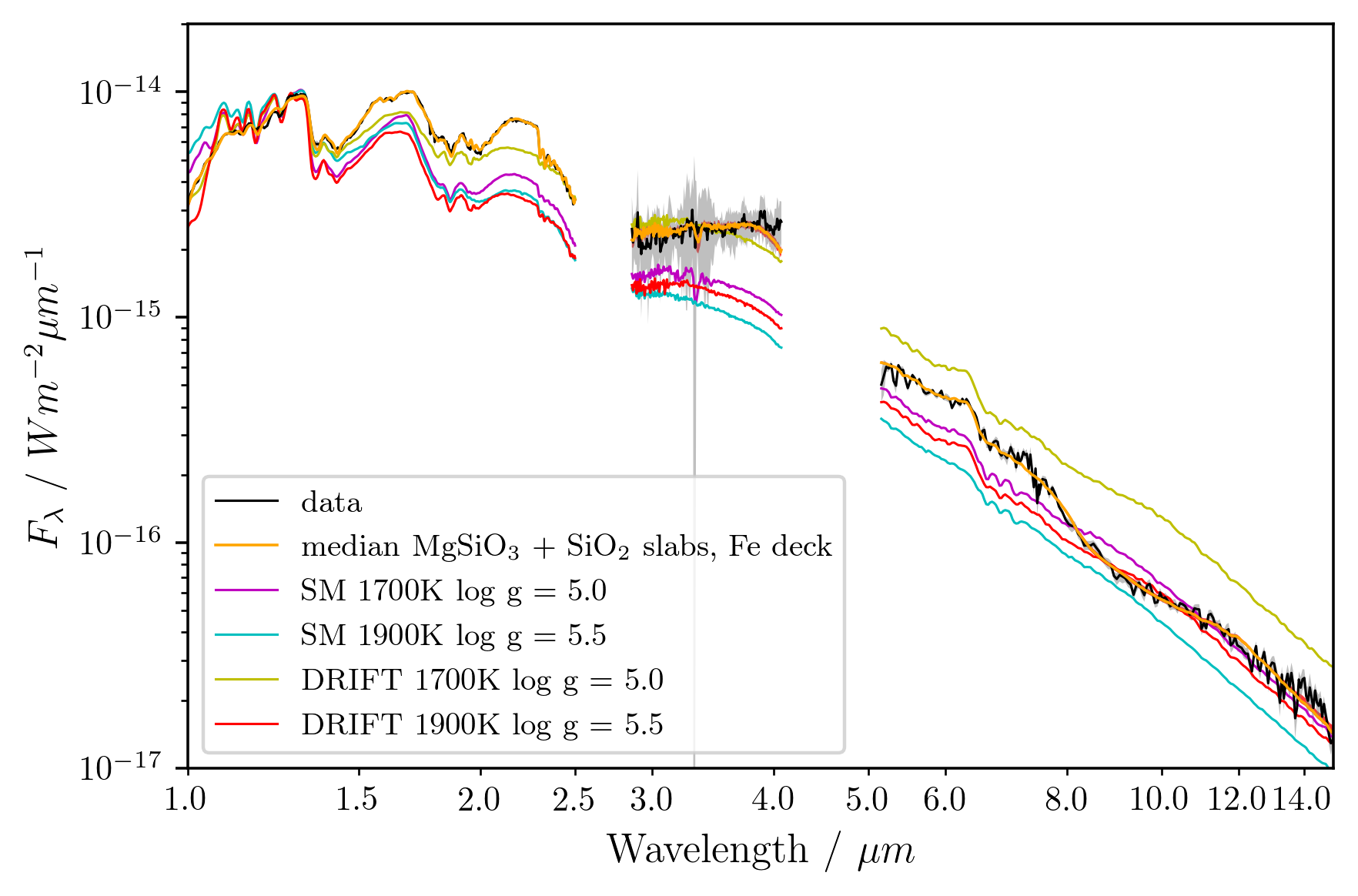}
\caption{Maximum likelihood retrieved model spectrum for the top-ranked model overlaid with the data. The grey shading indicates the uncertainty in the data. Pink shading indicates the 95\% confidence interval in the retrieval, and is mostly hidden by the line weight. Self-consistent grid models are shown for comparison,  and are  scaled to match the $J$~band flux in the observed spectrum.
\label{fig:med_fullspec}}
\end{figure*}

Figure~\ref{fig:med_fullspec} shows the model spectrum for the median set of parameters from the top-ranked model. 
This shows for the first time a model fit to the 1-15$\micron$ spectrum of an L dwarf that successfully fits the feature in $8 - 10 \mu$m region of the{\it Spitzer} IRS data, suggested to be due to enstatite clouds \citep{cushing2008,stephens2009}. 
In addition, our retrieval model is able to reproduce the full shape of the spectrum between 1 and 15$\mu$m.
With 36 parameters, presenting the properties of the preferred model is challenging in its own right. A ``corner" or ``staircase" plot such as is typically used to show correlations and degeneracies in the posterior distributions of multivariate model fits is impractical for all 41 parameters. For this reason we provide a full 41-parameter corner plot as an online plot for completeness, and here break the parameters into subgroups which are visualised separately and discussed separately.

\subsection{Thermal profile}
\label{sec:therm}

\begin{figure*}
\hspace{-0.8cm}
\includegraphics[width=300pt]{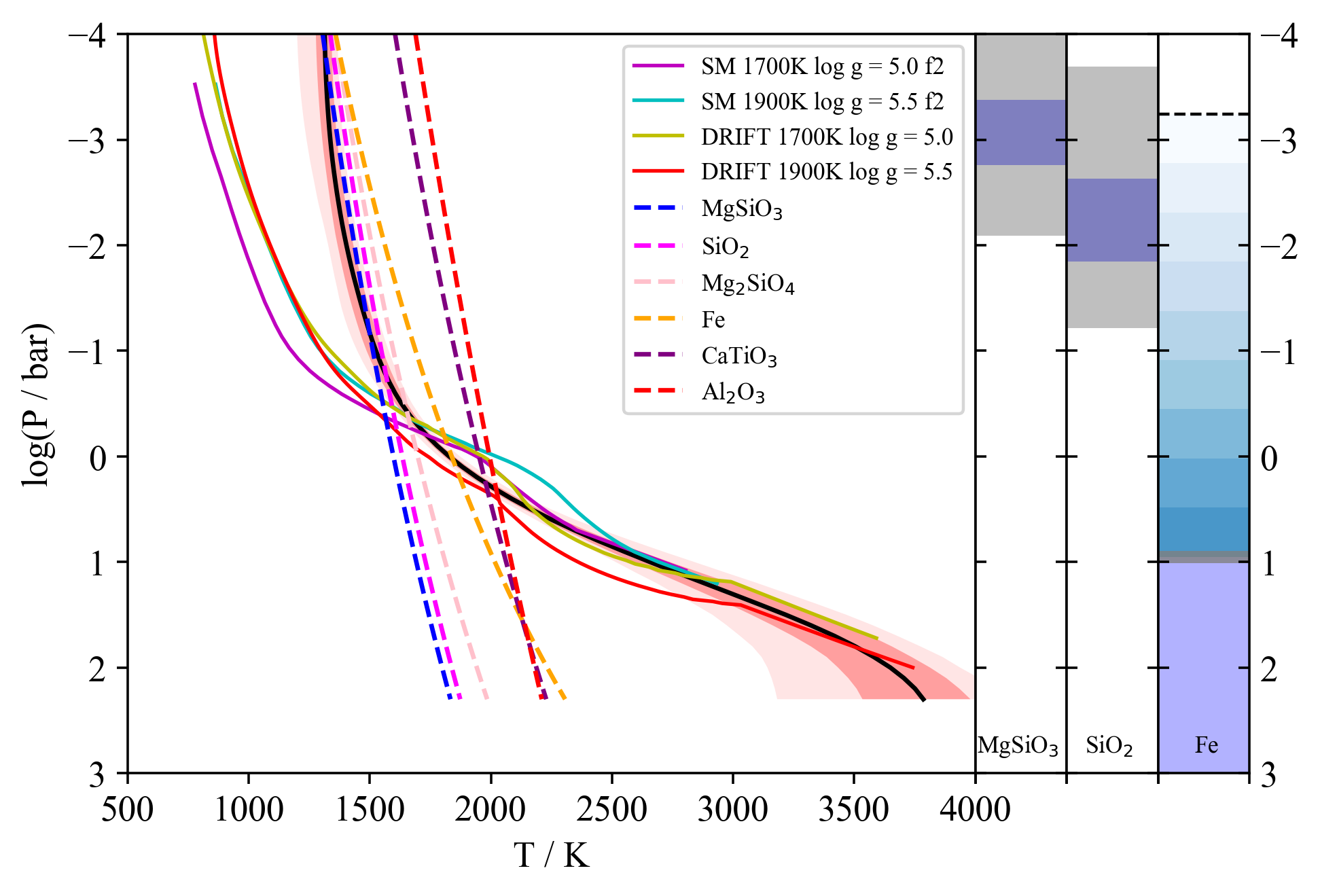}
\caption{Retrieved thermal profile (black line, pink shading for 1$\sigma$ and 2$\sigma$ intervals) and cloud pressures for 2M2224-0158, plotted with self-consistent model profiles from the Drift-Phoenix \citep[prefixed Drift, ][]{helling+2008} and Saumon \& Marley grids \citep[prefixed SM, ][]{sm08}. Also plotted are phase-equilibrium condensation curves for various potential condensates. The clouds pressures are indicated in bars to the left of the P/T profile. Blue shading indicates the median cloud location for the two slab cloud, with grey shading indicating the 1$\sigma$ range. The deck cloud is furthest to the right, with the uniform blue shading indicating the optically thick extent of the cloud deck, and grey shading the 1$\sigma$ interval for the $\tau = 1$ location. Graduated shading shows the range over which the deck cloud optical depth drops to 0.5.
\label{fig:profile}}
\end{figure*}

Figure~\ref{fig:profile} shows the retrieved thermal profile along with pressure depths for the clouds, and comparison to self-consistent grid models and phase-equilibrium condensation curves. 
The comparison model profiles are for \teff = 1700~K and 1900~K, $\log g = 5.0$ and 5.5 atmospheres taken from the DRIFT-Pheonix \citep{helling+2008} and \citet{saumon08} $f_{sed} = 2$ model grids. 
The self-consistent grid models have been selected to bracket both our extrapolated $T_{\rm eff}$ and $\log g$, and those estimated from the bolometric luminosity of the target and radii taken from evolutionary models \citep{filippazzo2015}. 
The deep thermal profile ($P \gtsimeq$~ 1 - 10~bar) follows the gradient of the comparison model profiles well. 
This in contrast to what was found when using only $1 - 2.5 \mu$m near-infrared data in B17, where there was a significant offset between the retrieved thermal profile and the grid model profiles at nearly all pressures. 
This highlights the importance of using as broad a wavelength coverage as possible when pursuing retrieval analysis on cloudy atmospheres.

Several noticeable differences are in the two relatively sharp gradient changes seen in three of the grid models, which are not seen in our retrieved profile. 
For example, in the \citet{saumon08} 1900K grid model profile, the gradient strays from the convective adiabat rising from deep in the atmosphere as thermal transport becomes radiation dominated at around 2500K, $\log P \approx 1.0$. 
The gradient shifts again close to $\log P \approx 0.0$ due to the formation of a detached convective zone driven by cloud opacity and the effect of the peak of the local Planck function overlapping with high gas opacity. 
This gives way to radiative transport again towards shallower pressures.
Our five parameter profile model is unable to capture such detail, so this difference cannot be interpreted as significant. 
However, since these features depend on details of how and where different opacities, such as clouds, come into play \citep[see e.g. ][]{marley2015} it will be an exciting prospect to develop methods for investigating such features in the coming years.

At shallow pressures we can see significant differences between our retrieved profile and the comparison profiles.
Specifically, our retrieved profile is considerably warmer at pressures $\ltsimeq 0.1$~bar, with an offset of nearly 500K from the warmest grid models at the top of the atmosphere. 
This is a similar result to that found in B17, and will be discussed further in Section~\ref{sec:strato}.

\subsection{Cloud properties}
\label{sec:cloudprops}

The retrieved parameters for our ``winning" model are summarised in Table~\ref{tab:cloud}. Our preferred model frames the cloud opacity as arising from three condensate species: ${\rm MgSiO_3, SiO_2}$ and ${\rm Fe}$. The first two of these are parameterised in slab clouds, while the Fe is parameterised as a deck cloud (see Section~\ref{sec:clouds}). 
The median locations of these ``clouds" in the final 2000 iterations of our \textsc{emcee} chain (1.1 million samples) are indicated by shaded bars to the right of Figure~\ref{fig:profile}. 

The two silicate clouds overlap in pressure location, and are found high in the atmosphere at pressures below 0.1 bar. 
The posterior distributions for their cloud bases show spreads of $\sim 1 {\rm dex}$ in $\log P$, reflecting relatively weak constraints on their depths in the atmosphere.  
The Fe cloud deck becomes optically thick at deeper pressures close to 10 bar. The location of the $\tau_{\rm Fe} = 1.0$ pressure level appears to be tightly constrained to $\log P {\rm (bar)} = 0.95 \pm 0.06 {\rm dex}$.  
The pressure change over which the optical depth of the deck cloud drops to 0.5 is constrained to $d \log P = 4.20^{+1.95}_{-2.25} {\rm dex}$, such that $\tau_{deck} = 0.5$ at $\log P = -3.25^{+2.25}_{-1.95}$. 
This thus represents a cloud opacity that is quite widely distributed through the atmosphere, despite the apparently tight constraints on its location.
This decay height is implemented in linear pressure space (see Equation~\ref{eqn:deck}, so this corresponds to a decay pressure scale of just under 10 bar, and thus $\tau_{deck} = 4.0$ at approximately 29 bar ($\log P = 1.46$).

\begin{table*}
\begin{tabular}{c c c c}
\hline
Cloud no. & 1 & 2 & 3 \\
\hline
Type & slab & slab & deck \\
Species & ${\rm MgSiO_3}$ & ${\rm SiO}$ & ${\rm Fe}$ \\
&&&\\
max $\tau_{cloud}$ at 1$\mu$m & $0.30^{+0.32}_{-0.18}$ & $4.54^{+0.58}_{-0.58}$ & N/A \\
&&&\\
reference pressure / $\log P {\rm (bar)}$ & $-2.76^{+0.66}_{-0.44}$ (max $\tau$)  & $-1.85^{+0.63}_{-0.99}$ (max $\tau$) & $0.95^{0.06}_{-0.06}$ ($\tau = 1.0$)\\
&&&\\
height / $d\log P$ & $0.62^{+0.67}_{-0.44}$ & $0.78^{+1.07}_{-0.58}$ & $4.2^{+1.95}_{-2.25}$\\
&&&\\
log(effective radius $a /\mu$m) & $-1.41^{+0.18}_{-0.17}$ & $-0.44^{+0.04}_{-0.20}$ & $-0.77^{+0.05}_{-0.06}$ \\
&&&\\
radius spread (Hansen distribution $b$) & $0.53^{+0.33}_{-0.36}$ & $0.03^{+0.18}_{-0.01}$ & $0.03^{+0.04}_{-0.01}$ \\
\hline
\end{tabular}
\caption{Summary of retrieved cloud properties for preferred model.
\label{tab:cloud}
}
\end{table*}

In addition to the profile and median cloud locations, Figure~\ref{fig:profile} also shows the condensations curves for several species, calculated for solar composition gas using the same methods as \citet{visscher2010a}.
Condensation curves such as these do not necessarily provide a prescription for what will condense, but instead provide a useful guide as to what \emph{can} condense.  
The pressure depths of our silicate clouds are both consistent with being located to the left of their respective condensation curves on our thermal profile. 
The $\tau = 1.0$ reference pressure for our deep Fe cloud deck corresponds to a temperature of of approximately 2500K. This is more puzzling, and will be discussed in more depth in Section~\ref{sec:clouddisc}.

Our top-ranked models all use Mie scattering clouds with a Hansen distribution of particle sizes, and models using either DHS scattering, or log-normal particle size distributions are very strongly rejected. 
Our top-ranked models also all include enstatite in its amorphous form, although our 6th-ranked model includes a mixture of amorphous and crystalline grains. 
All three retrieved clouds are dominated by sub-micron grains and have a negligible number of particles sized larger than $1\mu$m (see Table~\ref{tab:cloud}). 
Particle distributions dominated by small particles were a feature of the retrieved parameters for all the cloudy models tested.

To gauge the plausibility of our retrieved cloud properties and provide suitable outputs for comparison to physically motivated cloud models, we have estimated the 
amount of material required to produce the retrieved clouds in our best fitting model. The simplest approach for this is to use the retrieved total optical depth at 1$\mu$m and the particle size distribution parameters to estimate the column density for each cloud. 
We find a ${\rm MgSiO_3}$ column density of $5.4^{+1.1}_{-0.7}\times 10^{-4}$~g~cm$^{-2}$, or by number $3.2^{+0.7}_{-0.4} \times 10^{18}$~cm$^{-2}$. 
For ${\rm SiO_2}$ we find a column density of $1.4^{+0.6}_{-0.2} \times 10^{-4}$~g~cm$^{-2}$, or $1.4^{+0.6}_{-0.2} \times 10^{18}$~cm$^{-2}$.

If we take our median cloud locations at face value, we can estimate that such cloud masses would account for all of the oxygen in these atmospheric layers, and would require significant enhancement of magnesium ($\approx 5\times$) and silicon ($\approx 10\times$) compared to solar ratios. 
However, our retrieved cloud locations are not so precise, and are (at best) a very broad brushed 1-dimensional analogue to the dynamic clouds in our target's 3-dimensional rotating atmosphere.
As such, it is may be useful to consider these estimated cloud masses through comparison to the available material above their condensation curves on our retrieved thermal profile. 
In this view, the silicate clouds account for approximately 30\% of the available oxygen. 
If we (na\"ively) assume that there are no other sinks for magnesium or silicon and that the relative proportions of ${\rm MgSiO_3}$ and ${\rm SiO_2}$ reflect the proportions of Mg and Si, we can estimate the this requires $({\rm Mg} / {\rm Si}) = 0.69^{+0.06}_{-0.08}$.

\subsection{Bulk Properties}
\label{sec:bulk}

\begin{figure*}
\hspace{-0.8cm}
\includegraphics[width=550pt]{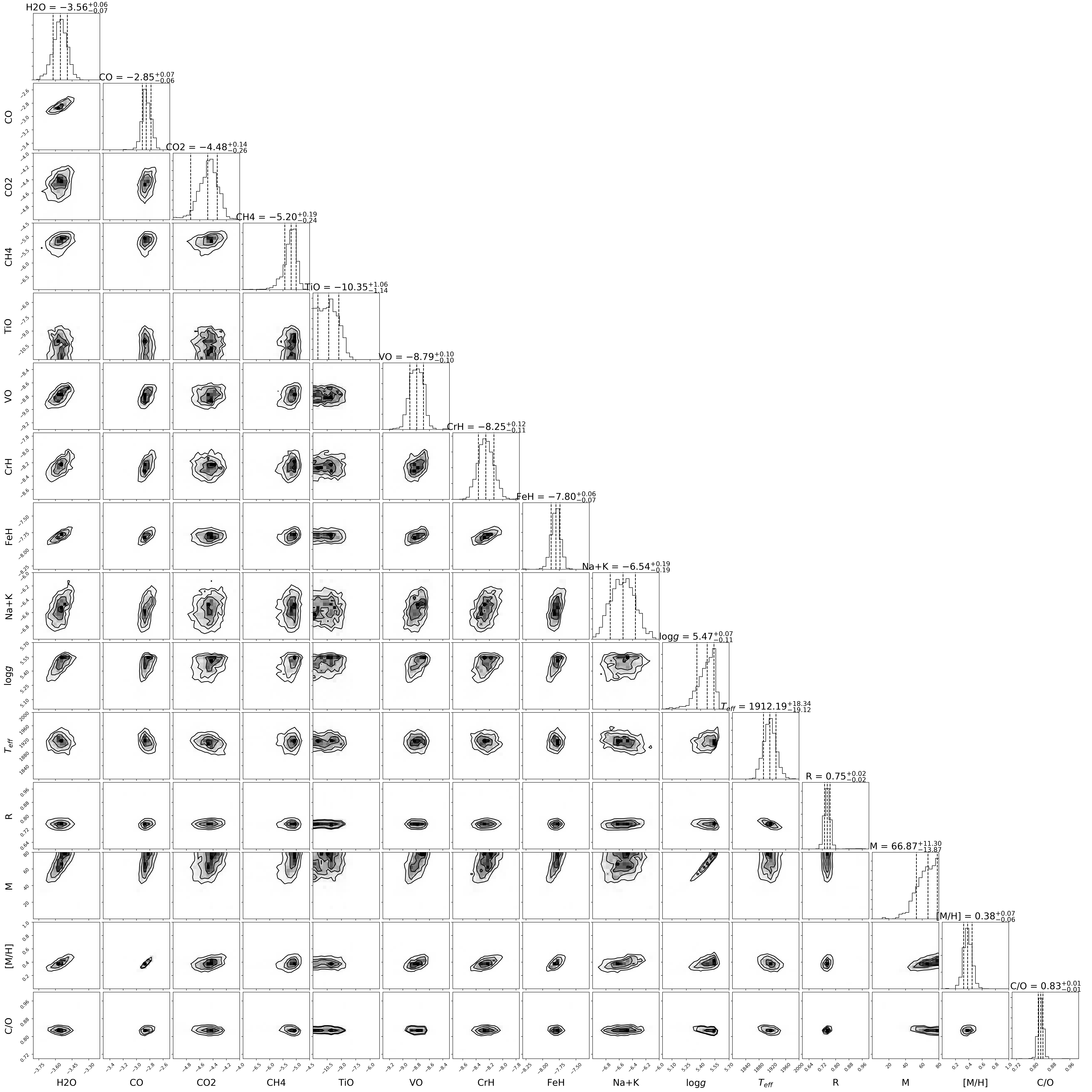}
\caption{Corner plot showing the top-ranked model retrieved parameters that set the gaseous opacity in the atmosphere, $\log g$, along with derived values for radius, mass, atmospheric metallicity and C/O ratio, and extrapolated $T_{\rm eff}$. See text for explanation of derived and extrapolated values. 
\label{fig:postcorner}}
\end{figure*}

Figure~\ref{fig:postcorner} shows the corner plot displaying posterior distributions for the retrieved gas fractions for absorbing gases in our model and $\log g$, along with derived values for radius, mass, atmospheric metallicity and C/O ratio, and extrapolated $T_{\rm eff}$. 
The derived and extrapolated values are found as follows. 
Radius is defined using the retrieved model scaling factor and Gaia parallax.
Mass is then found using the derived radius and the retrieved $\log g$,
$L_{\rm bol}$ is found by extrapolating the retrieval model to cover the $0.5 - 20$~$\mu$m range, summing the flux, and scaling it by $4\pi D^{2}$, where D is the distance defined by the Gaia parallax. $T_{\rm eff}$ is then found using the extrapolated $L_{\rm bol}$ and inferred radius.
The atmospheric C/O ratio is estimated by assuming all carbon and oxygen in the atmosphere exist within the absorbing gases considered in the retrieval. 
The metallicity is estimated by considering elements with our retrieved absorbing gases, and comparing their inferred abundances to their solar values from \citet{asplund2009}.

\subsubsection{Bolometric luminosity}
\label{sec:fund}
\begin{figure*}
\hspace{-0.8cm}
\includegraphics[width=500pt]{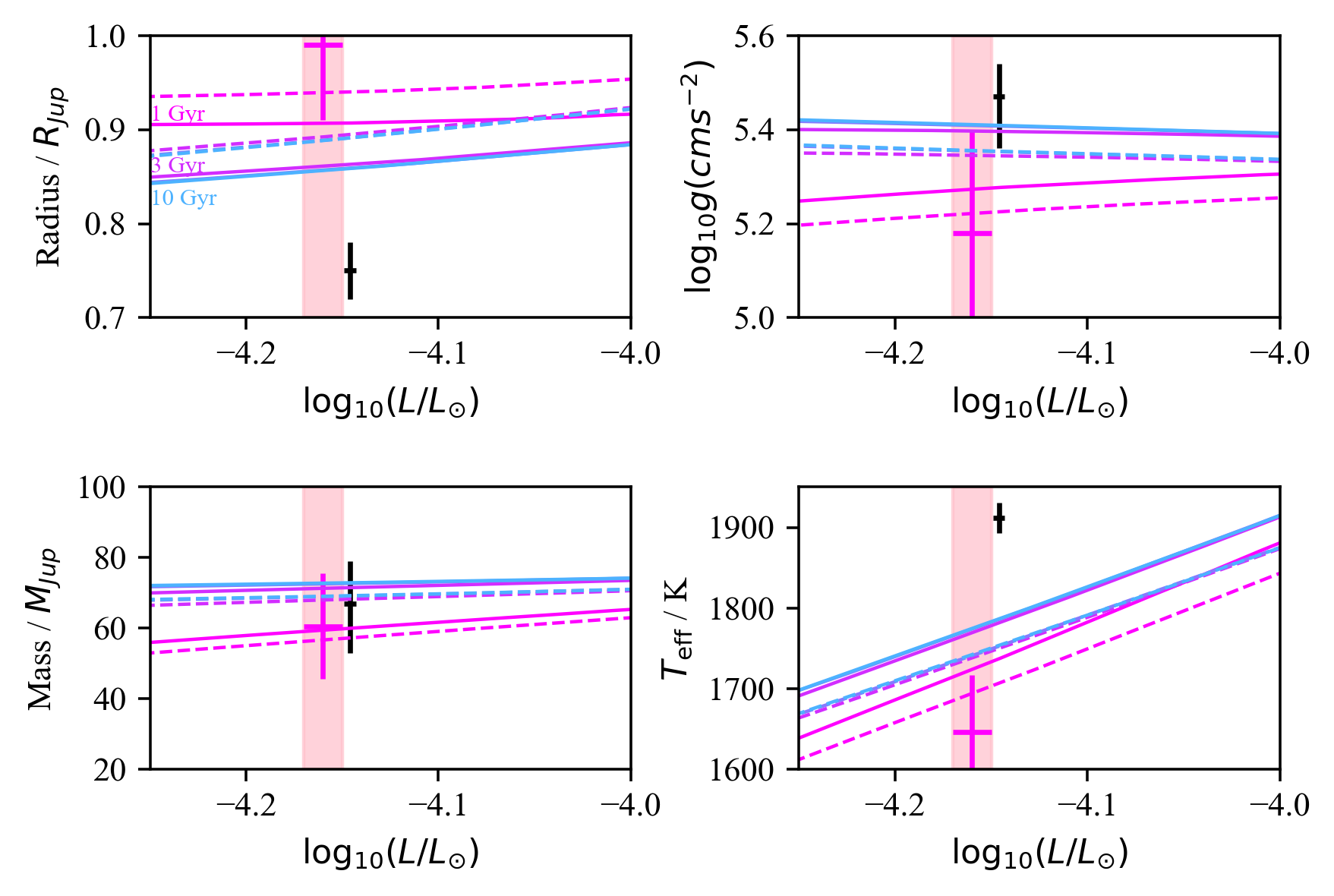}
\caption{Comparisons of our retrieved and extrapolated values for bulk properties of 2M2224-0158 with the empirical bolometric luminosity calculated by \citet[][;shaded pink]{filippazzo2015} and Sonora-Bobcat cloud-free evolutionary models \citep{marley2020} for solar metallicity (solid lines) and [M/H] = +0.5 (dashed lines).  Magenta error bars are the semi-empirical values from \citet{filippazzo2015} which used a combination of DUSTY00 \citep{chabrier2000}, SMHC08 \citep{sm08} and $f_{\rm sed} = 2$ \citet{sm08} isochrones. Our retrieval based estimated values are indicated with black error bars. Their derivation is described in Section~\ref{sec:bulk}. 
\label{fig:evocomp}}
\end{figure*}

Figure~\ref{fig:evocomp} shows a comparison of our retrieved and extrapolated values for bulk properties of 2M2224-0158 with the empirical bolometric luminosity calculated by \citet[][shaded pink in Figure~\ref{fig:evocomp}]{filippazzo2015} and Sonora-Bobcat cloud-free evolutionary models \citep[][; Marley et al in prep.]{marley2020}. 
Our extrapolated bolometric luminosity for 2M222-0158 is $\log (L_{\rm bol} / \Lsun) = -4.146 \pm 0.003$. 
This is similar to, but slightly higher than, the empirical value found by \citet{filippazzo2015} of $\log (L_{\rm bol} / \Lsun) = -4.16 \pm 0.01$.
Our derived radius, however, is considerably smaller than predicted by evolutionary models such as the Sonora-Bobcat grid plotted on Figure~\ref{fig:evocomp}, or the models used by \citet{filippazzo2015}, and this leads to a higher $T_{\rm eff}$.
This discrepancy will be further discussed in Section~\ref{sec:radius}.

\subsubsection{Composition}
\label{sec:comp}
Our retrieved gas abundances provide constraints on the composition of the photosphere of 2M2224-0158. 
Through consideration of the chemical, condensation and dynamic processes at work in its atmosphere, these in turn can provide useful insights to its bulk composition, as derived from its natal environment.  
Studies of brown dwarfs such as 2M2224-0158 can serve as essential due diligence for developing and validating such methodologies.

The most commonly deployed, and blunt, instrument for discussing composition in astronomy is aggregated metallicity [M/H], for which [Fe/H] is often used interchangeably as a proxy. 
Whilst our chemical equilibrium model allows us to directly access the metallicity of our target, our winning model requires us to estimate it based on the snapshots of photospheric abundance that our vertically constant gas fractions effectively provide. 
As discussed in the previous section, many of the species we are considering are expected to condense beneath and within our photosphere, as can be seen from the large decrease in atmospheric abundance of FeH, TiO and CrH with decreasing pressure in Figure~\ref{fig:chemeq}. 
Our retrieved gas fractions reflect these condensation processes, and would lead to a significantly sub-solar metallicity estimate if used in isolation. 
For example, based on our retrieved FeH gas fraction alone, we would estimate a value for ${\rm [Fe/H] = -2.2 \pm 0.06}$. 
These gases are, however, relatively minor contributors to our metallicity estimate, which is dominated by carbon and oxygen. 
As such they can be included or ignored with negligible impact on our metallicity estimate of ${\rm [M/H] = 0.38^{+0.07}_{-0.06}}$

 Of particular interest is the C/O ratio, which is viewed as a key diagnostic for investigating the formation environments and mechanisms for giant exoplanets \citep[e.g. ][]{oberg2011,nowak2020}. 
We estimate an atmospheric C/O ratio for 2M2224-0158 of $0.83 \pm 0.01$ based on the assumption that all carbon and oxygen are tied up within the considered absorbing gases. This atmospheric C/O ratio does not account for any oxygen that may be tied up in condensates such as ${\rm MgSiO_3}$ and ${\rm SiO_2}$. 
Our retrieved silicate cloud locations are at much shallower pressures than most of the gas opacity, $\ltsimeq 1\%$ of the atmosphere exists above the point where the ${\rm MgSiO_3}$ and ${\rm SiO_2}$ condensation curves cross our retrieved thermal profile. Any correction to the oxygen abundance due to clouds above this point will be small.

Our standard selection of gaseous absorbers in these experiments did not include SiO, which has very weak opacity and has not previously been identified in L dwarf spectra. 
However, predictions from thermochemical grids \citep[e.g.][]{visscher2010a} suggest that ${\rm SiO}$ should be a significant home for oxygen in L dwarf atmospheres after ${\rm CO}$ and ${\rm H_2O}$. 
If we account for oxygen tied up in ${\rm SiO}$ gas at the level predicted by our thermochemical equilibrium grid for the thermal profile of our winning model, we estimate that its inclusion might be expected to increase our oxygen abundance by about 8\%, reducing the C/O ratio to approximately 0.75.

To further investigate this, we have run an addition retrieval model based on our winning cloud model for which ${\rm SiO}$ opacity is included. In this case, the ${\rm SiO}$ abundance is only weakly constrained with $\log f_{SiO} = -3.59^{+0.33}_{-6.13}$, consistent with its small absorbing cross-section. This results in a C/O ratio of $0.83^{+0.06}_{-0.07}$. 
We thus adopt this wider uncertainty range for our C/O ratio.

Our uncorrected atmospheric C/O ratio is significantly super-solar, and is amongst the highest C/O ratios found in the stellar population \citep[e.g.][]{nissen2013,pavlenko+2019,stonkut+2020}. 
However, it has also been found that metallicity and C/O ratio are positively correlated due to oxygen being relatively less abundant in high-metallicity thin disk stars \cite{nissen+2014}. 
In this context, our retrieved C/O ratio and metallicity appear roughly consistent with one another.


\section{Discussion}
\label{sec:disc}

\subsection{Comparison to B17}
\label{sec:b17comp}
The future capabilities of {\it JWST} will offer wide and detailed wavelength coverage of a wide range of substellar and exoplanetary atmospheres. 
This work can provide data-driven insights into how additional wavelength coverage can impact retrieval studies of cloudy atmospheres in the temperature range considered here.

\begin{figure}
\hspace{-0.8cm}
\includegraphics[width=300pt]{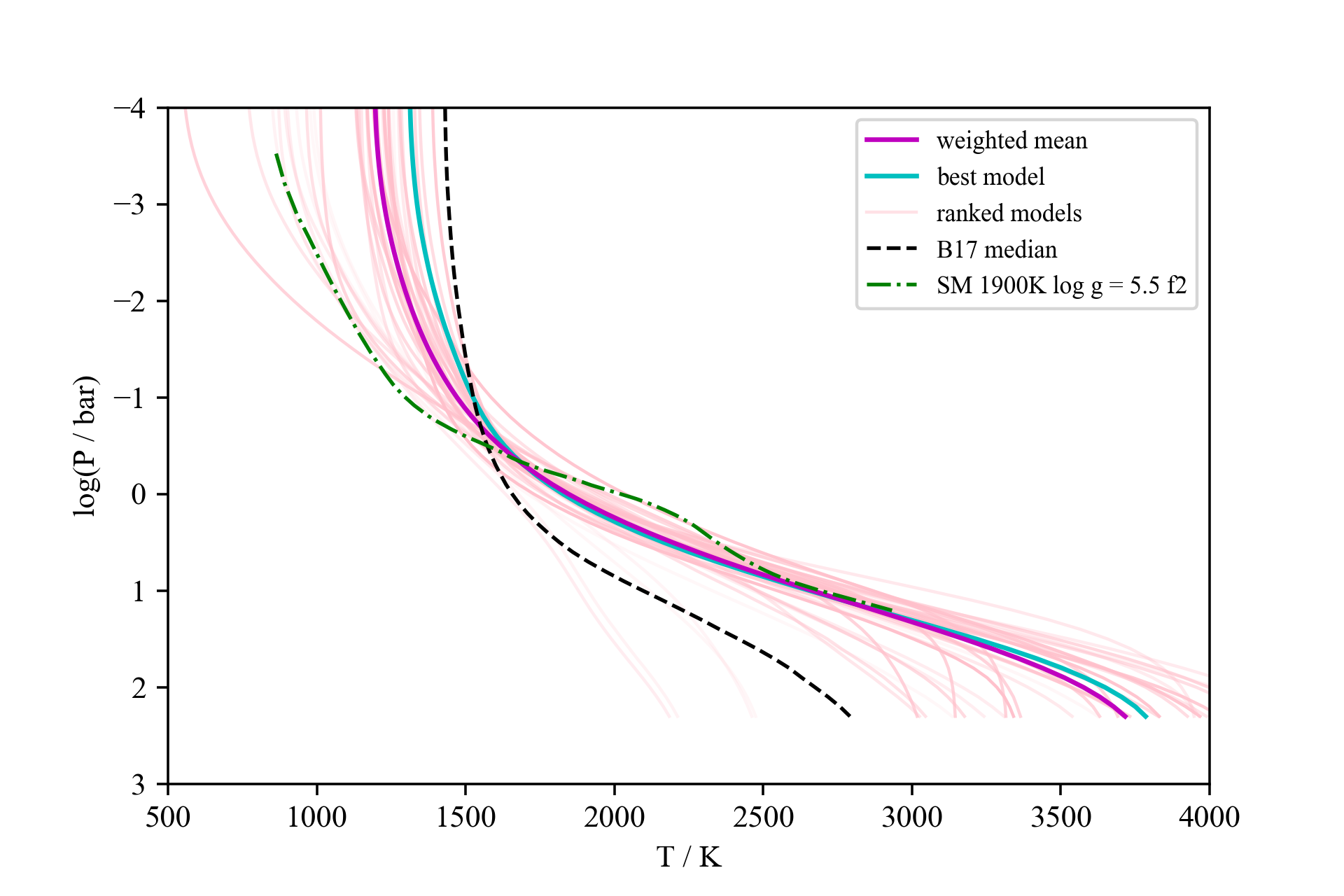}
\caption{A comparison of retrieved thermal profiles under different cloud model assumptions. 
Models other than our preferred model are plotted with line opacities proportional to ($1 / \log(\Delta BIC)$. The preferred model profile and the likelihood weighted mean of the profiles are indicated.
Also plotted are the median retrieved profile for 2M2224-0158 from B17, and an example self-consistent grid model that closely matches our retrieved $T_{\rm eff}$ and $\log g$. 
\label{fig:metaprof}}
\end{figure}

The most notable difference is in the shape of our retrieved thermal profile. 
Figure~\ref{fig:metaprof} compares our ``winning" model thermal profile to those of lower-ranked models covering a wide range of cloud assumptions, the retrieved profile from B17, and an example self-consistent grid model selected to match our retrieved parameters for 2M2224-0158. 
The difference between our winning model profile using the full 1 - 15$\mu$m range of available spectroscopy and that found using just $1 - 2.5 \mu$m Spex data is clear. 
The profile retrieved here provides a close match to the self-consistent grid model at pressures deeper than around 1 bar, whereas the B17 profile disagrees with the grid profile at all pressures.
This represents a shift of over 500~K at the bottom of our modelled atmosphere.  
Some of this difference may arise due to the more complex cloud model employed here. However, it is striking how similar the profiles for our highly ranked models are, despite differing cloud species, and structures. 
This is due to the fact that the large wavelength range allows for regions that have little cloud opacity to set the gaseous opacities and define the shape of the thermal profile. 
In B17 this region was much smaller (since we only covered $1 - 2.5~\mu$m), and a greater portion of the coverage was also impacted by problematic opacities such as the pressure broadened wings of the 7700\AA~\ion{K}{I} resonance line.
This highlights the benefit of broad wavelength coverage for remote sensing thermal structure in atmospheres, particularly in the absence of well understood cloud properties.

Along with a new thermal profile, we see a decrease of approximately 0.4~dex in our ${\rm H_2O}$ and CO gas fractions. 
Since these species dominate the C and O budget and each is shifted by a similar amount, our estimated C/O ratio is largely unchanged. 
This shift in gas fractions is likely linked to the new profile. 
For the most part it is a steeper temperature gradient through the photosphere, thus requiring less absorbing gas to achieve a similar reduction in emergent flux. 
Since ${\rm H_2O}$ and CO are the biggest contributors to gaseous opacity, it is not surprising to see them reduced. 
This reduction brings the overall estimated metallicity for this object much closer to the expected range for the Solar neighbourhood. 
This again highlights the importance of wide wavelength coverage to constrain the thermal profile and allow accurate abundance estimates.

We also see a significantly altered alkali abundance, with a -1.2~dex shift in our combined \ion{Na}{} and \ion{K}{} fraction, bringing it into conflict with the otherwise super-solar metallicity. 
The~\ion{K}{} opacities are particularly problematic due to the poorly constrained behaviour of the pressure broadened wing of the 7700\AA~\ion{K}{I} resonance line. 
This hinders our models' ability to simultaneously fit narrow \ion{K}{}~features and the shape of the pseudo-continuum around 1$\mu$m as set by red wing of the 7700\AA~\ion{K}{I} line, and also may impact our ability to retrieve accurate abundances.
Physical interpretation of this low abundance of alkali species would thus be premature.
This is an ongoing issue that that drove the decision to exclude wavelengths blue-ward of 1$\mu$m in this work, along with others \citep[e.g. ][]{line2015,line2017, burningham2017,gonzales2020}.
\citet{kitzmann2020} did not exclude this region and noted problems fitting the ~1$\mu$m $Y$~band peak in their study of T~dwarf. 
They used an updated description of the 7700\AA~\ion{K}{I} line wings from \citet{nallard2016}, suggesting that the challenge of accurate opacities in this region remains.

The retrieved radius is also significantly smaller than in B17, which was consistent with predictions of evolutionary models. This will be discussed in more detail in the Section~\ref{sec:radius}.

\subsection{Stratospheric heating}
\label{sec:strato}
At shallow pressures, our retrieved thermal profile is significantly warmer than the predictions of self-consistent grid models (see Figure~\ref{fig:profile}).
Moreover, Figure~\ref{fig:metaprof} demonstrates that this is common feature of our highly ranked models.
Although there is some scatter in the model profiles, the preference for temperatures above around 1200K is clear.
Our top-ranked thermal profile decreases from around 1600~K at $\log P {\rm (bar)}= -0.5$, to roughly 1300~K at $\log P{\rm (bar)} = -4.0$. 
This is in contrast to the self-consistent models, which exhibit temperatures falling from 1600~K to below 900~K in the same range (see Figure~\ref{fig:profile}).
This comparison is for cloudy solar composition self-consistent models, since we do not yet have access to cloudy models that span the metallicity and C/O ratio inferred for our target. 
However, cloud-free models show very small impact ($\sim 10$~K) on stratospheric temperatures due to changing composition. 
For a 1900~K, $\log g = 5.0$ model, an increase in metallicity to ${\rm [M/H] = 0.5}$ reduces the temperature at $\log P{\rm (bar)} = -4.0$ by 15~K compared to solar composition.
Increasing the C/O ratio to 0.9, raises the temperature in the same region by 10~K.

It is worth considering if this divergence from the self-consistent model predictions is perhaps some artefact of our adopted parameterisation of the thermal profile. 
In B17 we tested the retrieval framework using this parameterisation against simulated data created using thermal profiles from the self-consistent model grid plotted in Figure~\ref{fig:profile}, and did not identify any bias to higher temperatures in the upper atmosphere. 
We also tested a retrieval using this parameterisation against the benchmark T8 dwarf G570D, and found a retrieved profile consistent with grid models and the retrieval of \citet{line2015}.

We have also tested the parameterisation employed by \citet{line2015,line2017}. This uses a low-resolution 13-point thermal profile with spline interpolation to the full- resolution pressure scale, but penalizes the second derivative of the final curve in the retrieval. This has the effect of minimizing the jaggedness of the profile unless the benefit to improving the fit is significant. 
However, the L~dwarf spectrum is relatively low-contrast compared to that of the T8 dwarfs for which the method was devised. 
As we previously found in B17, the result is that the data do not justify anything other than an entirely linear profile, even with the inclusion of the long-wavelength data used here. 
A more in-depth exploration of alternative ways to parameterise the thermal profile is beyond the scope of this work. We can thus not rule out the possibility that alternate parameterisations may find different profiles that may also fit the data.

As we noted in B17, the warm retrieved stratospheric temperature is consistent with previous model fitting by \citet{sorahana2014}, who found evidence for a roughly isothermal profile, with T = 1445~K, at pressures shallower than $\log P {\rm (bar)} \ltsimeq -0.5$ for this object.  
These results thus tell a similar story for the upper atmosphere of this object, and suggest an energy transport mechanism that is not currently incorporated in self-consistent radiative-convective equilibrium models. 
The nature of this heating is uncertain, but it may be common amongst L~dwarfs \citep[][B17]{sorahana2014}. 
The detection of chromospheric activity via H$\alpha$ emission in several objects with apparent shallow-pressure heating (including 2M2224-0158) lead \citet{sorahana2014} to argue that the mechanism is magnetic-hydrodynamic in nature, and it is understood in the context of stellar chromospheric heating. 
The apparent presence of cloud layers in the heated region, however, is more reminiscent of the unexpectedly warm stratospheres and thermospheres of Solar System gas planets \citep[e.g.][]{appleby1986,seiff1997,MarleyMcKay1999}.  
These heated regions are not thought to be due to auroral or irradiation, but are currently understood in terms of gravity waves that propagate from deeper in the atmosphere before breaking in the stratosphere and depositing their energy \citep{schubert2003,freytag2010,ODonoghue2016}.

\subsection{Radius}
\label{sec:radius}

The radius implied by our top ranked retrieval model is significantly smaller than predicted by evolutionary models (see Figure~\ref{fig:evocomp}. 
This is a surprising, and difficult to interpret result which is a significant departure compared to our previous study of 2M2224-0158 in B17, which found a radius that was consistent with evolutionary models.

An examination of the retrieved radii across our ensemble of models finds that our winning model is towards the smaller end of the distribution, which has a median of 0.82~\Rjup, and 16th- and 84th-percentiles of 0.76~\Rjup~and 0.97~\Rjup~respectively. 
The upper end of this distribution is consistent with the predictions of various evolutionary models. 
However, the retrieval models with the larger radii are some of the most poorly ranked models in our set, with $\Delta BIC \gtsimeq 100$.

To further investigate this issue we have rerun our winning model with a Gaussian prior on the radius with mean of 0.82~\Rjup, $\sigma = 0.04$~\Rjup. 
The retrieved parameter set was nearly identical to that retrieved with a 0.5-2.0~\Rjup~uniform prior on the radius, with all values consistent within $1\sigma$, and most much closer than that. The retrieved radius with the Gaussian prior was $0.77 \pm .02$~\Rjup, compared to $0.75 \pm 0.02$~\Rjup~with a flat prior.

Rerunning all of our models with the Gaussian prior is computationally prohibitive. 
However, we can nonetheless consider whether this Gaussian prior would be likely to impact our model selection, and rank a model that retrieved a radius closer to the expected value more highly than our winning model.
To do so we must assume that the Gaussian radius prior will have similarly small effects on retrieved parameters of the other models, and that their fits to the data are similarly good/bad as with the uniform prior. 
The large deviation from our expected radius for our winning model was penalised in the likelihood, and increased BIC value by 16. 
Models with radii closer to the center of the Gaussian prior will be penalised to a lesser degree, and so will have smaller increases to their BIC values. 
However, since none of the large radius models have $\Delta BIC < 100$, we can conclude that introducing a Gaussian prior would have a minimal impact on our model selection results.

Elsewhere, a similarly small radius was found by \citet{gonzales2020} for the low-metallicity s/dL7 dwarf SDSS~J1416+1348A, also using the {\it Brewster} framework. Using the Helios.R-2 framework, \citet{kitzmann2020} found a smaller than expected radius for the T1 dwarf $\epsilon$ Indi Ba under both chemical equilibrium and vertically constant composition assumptions, and for the T6 dwarf $\epsilon$ Indi Bb under the vertically constant composition assumption.

By contrast, retrieved radii for very late type T dwarfs have generally been found to be in agreement with evolutionary models. For example, the T7.5p companion SDSS~J1416+1348B was found by \citet{gonzales2020} to have retrieval radius consistent with low-metallicity Sonora-Bobcat models using {\it Brewster}. 
Using the {\tiny CHIMERA} package, \citet{line2017} found retrieved radii for 9 out of 11 T7 and T8 dwarfs were consistent with the evolutionary model estimates used in \citet{filippazzo2015}.

\citet{sorahana2013} also found smaller than predicted radii for L~dwarfs and consistent radii for T~dwarfs via fitting of self-consistent models to AKARI spectroscopy. 
This suggests that this radius problem is not an artefact of some bias in data-driven retrieval frameworks.

\citet{kitzmann2020} have suggested that the small retrieved radii in their study of $\epsilon$~Indi~BaBb could arise from a heterogeneous atmosphere, presumably with dark regions that contribute little flux and resulting in a smaller effective emitting area. 
Such a scenario is consistent with the apparent lack of a radius problem in late-T dwarfs that are thought to have essentially cloudless photospheres. 
In the case of 2M2224-0158, it would require the equivalent of some 30\% of the top of the atmosphere contributing zero flux to the spectrum to account for our difference with a typical field age evolutionary model. 
In the cases studied in \citet{kitzmann2020}, the equivalent zero flux covering area would be much larger, as their inferred radii are as small as 0.5\Rjup.
Such a substantial covering fraction might be expected to produce some signal of rotational variability, as seen across so much of the brown dwarf population. 
However, neither $\epsilon$~Indi~BaBb, SDSS~J1416+1148A, nor 2M2224-0158 have convincing detections of rotational modulation \citep[][Vos private communication]{hitchcock2020,khandrika2013,metchev2015,milespaez2017}. 
If dark patches are indeed present, this suggests that either these objects are presenting with unfavourable geometry for detecting rotational signals, or the dark regions are arranged latitudinally in bands.

If such dark regions correspond to differences in cloud cover, then this cloud cover must be quite different to that suggested by our retrieval results. 
Most of our wavelength region is not strongly affected by the cloud opacity we have retrieved, so the additional cloud cover would need to be essentially grey by comparison and lying higher than around 0.1 bar, in order to effectively reduce the total flux without significantly altering the shape of the SED. 
This would require much larger particles than those implied by our retrieval results thus far.

It is also possible that our retrieved radius is correct, and that the shortcoming arises in the evolutionary models. 
The Transiting Exoplanet Survey Satellite (TESS) has revealed several new transiting brown dwarfs in the $40 - 70$~\Mjup~range with ages in the $3 - 6$~Gyr range which have radii close to, or smaller, than that predicted for 10~Gyr old objects \citep{carmichael2020a,carmichael2020b}. 
\citet{carmichael2020b} speculate that this could be accounted for by the brown dwarfs reaching their asymptotic minimum radius of 0.75~\Rjup~by the age of 4 or 5~Gyr, rather than 10~Gyr as predicted by most evolutionary models.
So far, this has only been seen for objects above about 40~\Mjup. If it is indeed the case that only higher mass brown dwarfs are affected then this would also give rise to the difference in results between L~and late T~dwarfs seen via spectroscopy, as at field ages these correspond to higher and lower mass populations. As the sample of transiting brown dwarfs grows the origin of this issue will become clearer.

\subsection{Cloud composition and location}
\label{sec:clouddisc}
Our investigation of the clouds in 2M2224-0158 explored a wide range of possibilities to make our study as data-driven as possible.
Our parameterisation of cloud opacity is intentionally independent of models of cloud condensation so that we can cleanly assess what input the data alone can make to said modelling efforts.
We used this flexibility to try a wide range of condensates including those predicted by cloud models, and some which are not expected but whose optical properties had the potential to match features in the data. 
Our preferred model combines a deep iron cloud with enstatite and quartz clouds at lower pressures. 
Whilst other condensates are likely present, this result suggests that these are the condensates that are dominating the cloud opacity in the photosphere.
These results provide a powerful test for different models of cloud formation and condensate chemistry.
A notable area of divergence between differing model approaches is the compositions of the predicted clouds.

A striking aspect of our preferred model is the presence of quartz (${\rm SiO_2}$), with abundance comparable to enstatite.
\citet{Helling2006} made a clear prediction for the presence of quartz in brown dwarf atmospheres as a dominant component of the silicate clouds.
This was in contrast to the predictions of phase equilibrium modelling that suggested that quartz should not exist in substellar atmospheres. 
\citet{visscher2010a} found that enstatite (${\rm Mg Si O_3}$) removes silicon from the gas phase so efficiently, that quartz can only condense if enstatite formation is suppressed somehow i.e. they found that quartz could not form alongside enstatite.

These predictions were made on the assumption of solar composition for which ${\rm Mg / Si} \approx 1$. 
In fact, the same mass balance and phase equilibrium considerations suggest that quartz is expected to co-exist alongside enstatite if silicon is more abundant than magnesium, i.e for atmospheres with an Mg/Si ratio of less than 1.
In Section~\ref{sec:cloudprops}, we estimated ${\rm Mg/Si} = 0.69^{+0.06}_{-0.08}$, assuming no other sinks for Mg and Si. 
${\rm Mg/Si}$ has been found to be negatively correlated with metallicity, and our estimate is comparable to outlier values seen for high-metallicity outliers in \citet{adibekyan2015} and \citet{suarezandres+2018}.

Another notable feature of our preferred cloud model is the absence of forsterite (${\rm Mg_2 SiO_4}$). 
The presence of forsterite is predicted by both microphysical and phase equilibrium models at solar composition \citep[e.g. ][]{lodders2006,Helling2006,visscher2010,gao2020}. 
Our model set includes a 4 cloud model comprising quartz, enstatite and forsterite slabs, along with a deeper iron cloud. 
In this case, the retrieved forsterite cloud was was either pushed out of view into the deeper atmosphere, or found at lower pressure with very low-optical depth i.e. it did not contribute to the emergent spectrum, and the model was down-ranked due to additional parameters.

To investigate whether forsterite is expected in this case, we have performed preliminary calculations of phase equilibrium cloud masses for our estimated values of ${\rm [M/H]}$, C/O and Mg/Si following \citet{visscher2010a}. 
We find that forsterite is expected to condense over a narrower temperature range than at solar composition (at 1 bar): $1510 < T < 1668$~K for our composition, versus $T < 1702$~K for solar. 
It is also predicted to be less abundant than enstatite, but more abundant than quartz. 
These preliminary calculations appear to present a conflict between our retrieved cloud properties and the predictions of phase equilibrium, and highlight the importance of exploring non-solar elemental abundances and/or disequilibrium effects in future cloud models.
\\

The presence of a deeper iron cloud is consistent with the predictions of both phase equilibrium and microphysical models discussed above. 
However, the reference pressure at which the iron deck cloud reaches $\tau = 1.0$ at 1$\mu$m is slightly shallower than 10 bar, and corresponds to a temperature of approximately 2500K on our retrieved thermal profile - too hot to match any of the condensation curves plotted on Figure~\ref{fig:profile}, including iron.
We found that this T-P point was also fairly consistent amongst the lower ranked models regardless of the compositions of shallower cloud species.

There are several things to consider in understanding this apparently problematic result. 
The wavelength dependence of the iron cloud opacity, and the impact of gaseous opacities, means that this deep cloud deck's impact on the spectrum is greatest near 1$\mu$m (See Figure ~\ref{fig:contfunc}).
As such, the retrieved reference pressure may be affected by issues with the problematic pressure broadened wing of the 7700~\AA~\ion{K}{I} resonance line.
It is worth noting the deeper reference pressure for \ion{Fe}{} cloud found here compared the cloud deck in B17 occurs alongside a reduction in the \ion{K}{} abundance.
Additionally, the Fe cloud opacity builds up through a wide of range of pressures, including regions to the left of the iron condensation curve. 
So, it is possible that the opacity is arising from several species, with iron producing the dominant spectral signature, and this could account for some of the extension beyond the iron condensation region. 

Somewhat speculatively, we could also imagine that deeper opacity could arise from molten iron raining out of the atmosphere and falling to deeper pressures before being vaporised. Opacity due to falling, and vaporising, rain can be observed on Earth in the form of {\it virga}, and such a phenomena taking place in this atmosphere would explain opacity arising beyond the condensation curve. 

\subsection{Cloud particle properties}

Our retrieval has also placed constraints on the particle size distribution for our modelled clouds, as well as testing alternative assumptions for modelling the scattering from our cloud particles. 
All of our highly ranked models share key features with our top model: a Hansen distribution dominated by sub-micron grains, whose optical properties are modelled according to Mie theory rather than as a distribution of hollow spheres \citep[DHS, ][]{min2005}.
\citet{Helling2006} suggested that DHS scattering should be considered for modelling brown dwarf cloud opacity to account for inhomogeneous particle shapes. 
Our retrieval's preference for Mie scattering appears to rule this out.
The preference for a Hansen distribution of particle sizes is consistent with the findings of \citet{hiranaka2016}, and contrasts with the popular log-normal size distribution used some self-consistent models \citep[e.g.][]{ackerman2001,sm08} and retrievals \citep[e.g.][]{nowak2020}.

A significant difference between our particle size distribution and those of nearly all equilibrium and microphysical cloud models is that ours is uniform with altitude in the atmosphere. 
This presents a significant hurdle when comparing our results with cloud model predictions. 
However, some useful comparisons can be made when we consider predicted particle sizes in the context of the atmospheric levels which contribute most to the shape of the emergent spectrum, as it is these regions that will drive the retrieval estimates for particle sizes.

\begin{figure*}
\hspace{-0.8cm}
\includegraphics[width=500pt]{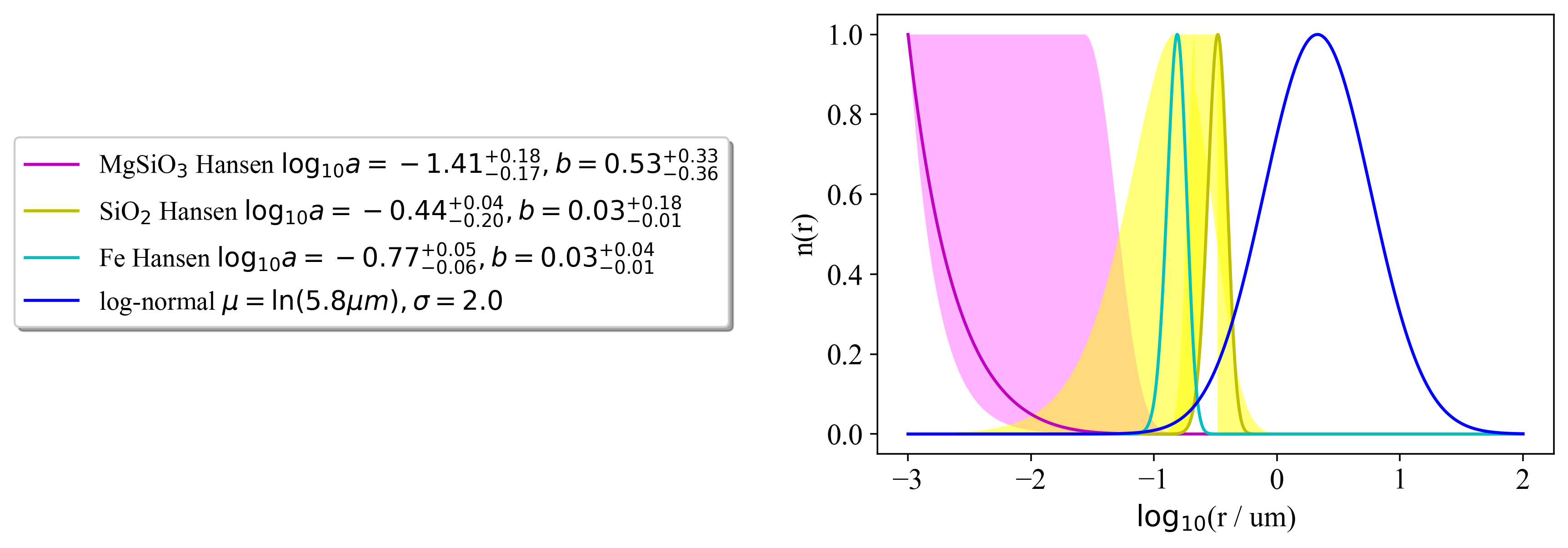}
\caption{A comparison of particle size distributions (normalised to have maximum values of 1) for our three retrieved cloud species, and the log-normal particle size distribution used in the \citet{saumon08} self-consistent model grid. The coloured shading indicates the range of the 16$^{th}$ and 84$^{th}$ percentiles of the retrieved size distributions. 
\label{fig:partsize}}
\end{figure*}

Figure~\ref{fig:partsize} compares our retrieved particle size distributions to a log-normal distribution with parameters $\mu = \ln(5.8 \mu m)$ and $\sigma = 2$. 
These parameters are as predicted for enstatite by the {\sc eddysed} cloud model \citep{ackerman2001} used in the Saumon \& Marley model grids, for the $\tau = 2/3$ level in a $\teff~=~1700$~K brown dwarf.  
All three retrieved clouds are dominated by much smaller grains than those predicted by the {\sc eddysed} model. 
In the case of the silicate grains, this difference is what allows our retrieved model to fit the broad spectral feature in the 9~-~10~$\mu$m region. 
The Mie scattering features are only apparent for relatively narrow distributions of small particles, whereas the broad log-normal distribution, incorporating large particles, smears out these Mie features. 

Figure~\ref{fig:partsize} also shows that for the enstatite cloud, with a large value for the Hansen b parameter (width), the distribution tends towards a power law. 
This is a similar functional form to the interstellar grain size distribution employed by the {\sc cond}, {\sc dusty} and {\sc Settl} models.

A common feature of cloud models such as EddySed and microphysical cloud models is a tendency to predict larger particle sizes deeper in the atmosphere and smaller particles at shallower pressures. 

Work by \citet{gao2018} using the CARMA microphysical cloud model found multi-modal size distributions were predicted for ${\rm K Cl}$ clouds in cool ($T_{\rm eff} \approx 400$K) substellar atmospheres, with the main peak in the distribution for very small ($\sim 10^{-3} \mu$m) particles found alongside a significant peak in the abundance of larger particles ($\sim 1 \mu$m). 
\citet{powell2018} also found multi-modal distributions for silicate clouds in hot-Jupiter atmospheres using the same modelling framework. 
In that case large ($< 1\mu$m) sized grains were predicted alongside sub-micron particles. 
Whilst neither of these predictions are directly applicable to the case studied here, they highlight the possibility of further complexity that could be missed by a retrieval model that assumes a single particle size distribution, which is uniform with altitude.

We also pursued a limited investigation into the presence of crystalline versus amorphous silicate grains. 
This was driven by the presence of apparently sharp features and structure in the spectral feature attributed to silicate clouds around 9$\mu$m, which would be consistent with crystalline grains. 
Our highly ranked models all feature amorphous grains of enstatite, which is consistent with the predictions from microphysical modelling by \citet{Helling2006}.

\subsection{Clouds and C/O ratios}
\label{sec:coclouds}

Given the ubiquity of clouds and the aforementioned interest in measuring C/O ratios of substellar objects and giant exoplanets, we now devote some discussion to considering how the presence of the former can impact the latter.

The impact of condensation processes on the atmospheric C/O ratio should be considered carefully. 
\citet{line2015,line2017} applied corrections to their retrieved C/O ratios for late-T dwarfs on the assumption of 3.28 oxygen atoms per silicon atom being removed from the atmosphere through condensation in ${\rm MgSiO_3}$ and ${\rm Mg_2SiO_4}$ \citet{burrows1999}. 
This translates as a roughly $25\%$ addition to the retrieved atmospheric oxygen abundance to be included in the C/O ratio calculation. 
It should be noted that this $25\%$ correction arises by assuming a solar ${\rm Si / O}$ ratio, and that this correction may thus need fine-tuning to account for a) different dominant condensates to those assumed for above; and b) non-solar abundance ratios.

We have detected ${\rm MgSiO_3}$ and ${\rm SiO}_2$ clouds in the upper atmosphere of 2M2224-0158. 
As discussed in Section~\ref{sec:cloudprops}, our median cloud masses account for around $30\%$ of the oxygen in the atmosphere above the condensation pressures for the detected species.
However, as these oxygen bearing clouds are at much shallower pressures than most of the gas opacity, any correction to the oxygen abundance due to clouds above this point will be small. 
A straightforward scaling of the oxygen sequestered in the detected clouds to account for the $\ltsimeq 1\%$ of the observable photosphere that they occupy reduces the correction to around $0.3\%$.
We thus neglected a cloud correction for the C/O ratio in this L dwarf. 
For cooler objects, such as late-T dwarfs, for which the entire photosphere lies above the condensation zone for these species we would expect the correction to correspond directly to the $30\%$ oxygen estimated to be taken up by silicate clouds in Section~\ref{sec:cloudprops}. 
This is comparable to the $25\%$ condensation correction estimated by \citet{line2015} for late-T dwarfs based on equilibrium cloud condensation expectations outlined in \citet{burrows1999}.

\begin{figure}
\hspace{-1cm}
\includegraphics[width=280pt]{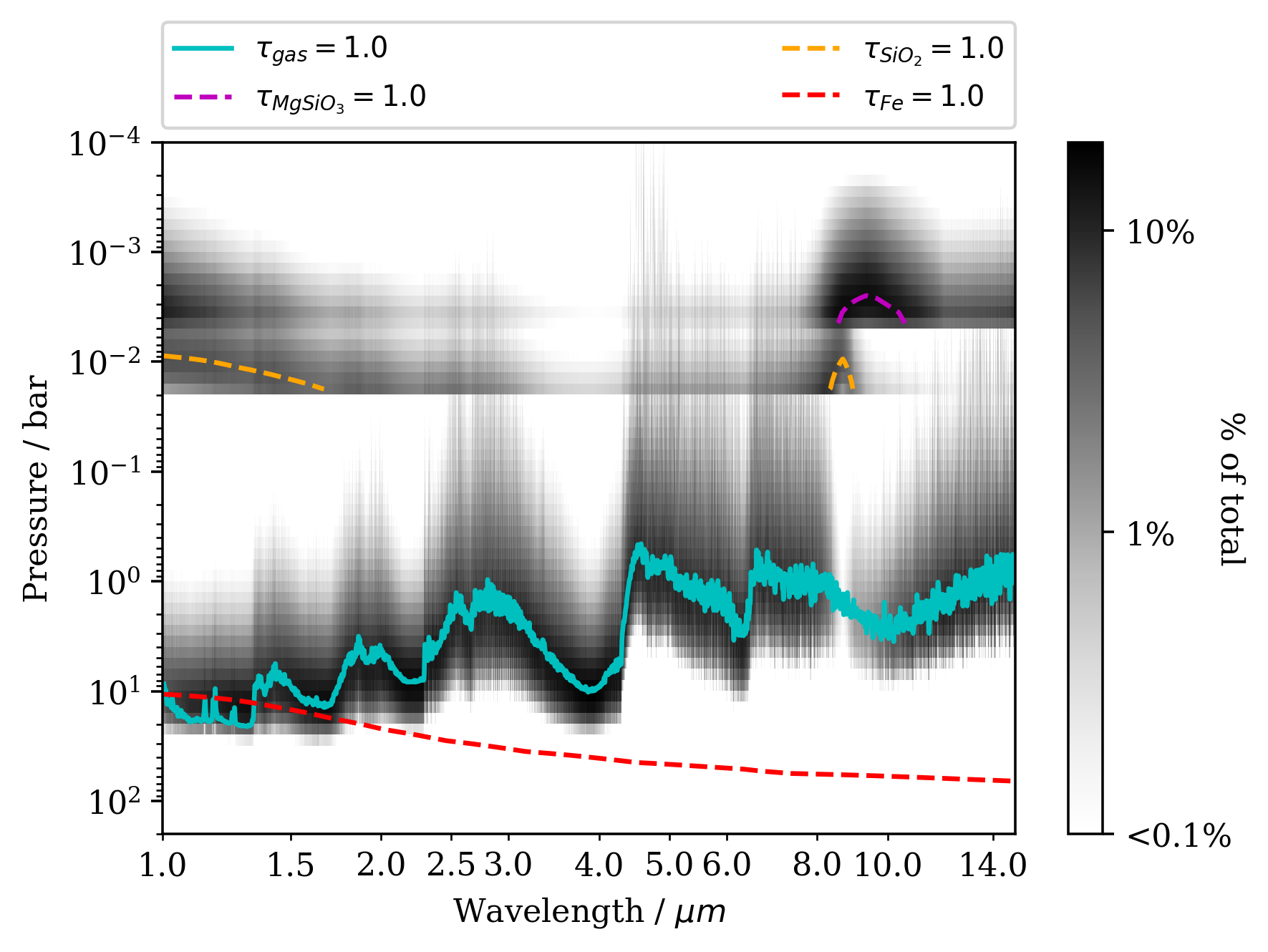}
\caption{Contribution function for a spectrum based on the maximum likelihood retrieved parameters for the top-ranked model. The contribution function in an atmospheric layer, lying between pressures $P_1$ and $P_2$ is defined as
$C(\lambda, P) =\frac{B(\lambda,T(P))\int_{P_1}^{P_2}d\tau}{\exp{\int_0^{P_2}d\tau}}$. $\tau = 1$ lines are included for gas phase opacities and each of our cloud species. 
\label{fig:contfunc}}
\end{figure}

Another point to consider is what impact the choice of cloud model has on the estimated C/O ratio. 
In this work, it appears it has surprisingly little impact on our estimated C/O ratio.
Most of our retrieval estimates for the C/O ratio lie quite close to that of our preferred model, across a wide range of cloud models, including cloud free, and hence different thermal structures. 
Taking the mean C/O ratio of all our model runs, we find ${\rm C/O} = 0.85 \pm 0.03$, compared to $0.83 \pm 0.01$ for the preferred case. 
As can be seen from Figure~\ref{fig:contfunc}, the bulk of our retrieval model's cloud opacity lies beneath the gas opacity for most of the spectral range considered here. 
A similar result was found by \citet{molliere2020} for the directly imaged exoplanet HR8799e. 
In that case they found that their retrieved C/O ratio was consistent between two different cloud models that resulted in quite different thermal structures. 

\subsection{Chemistry}
\label{sec:chem}
A surprising outcome of this work is that the preferred model is one which assumes vertically constant gas fractions, as opposed to one which assumes thermochemical equilibrium abundances of absorbing gases. 
All available \teff estimates for our target, whether based on bolometric luminosity or extrapolated from our retrieval, suggest that it lies in the regime where the chemical timescales in its photosphere are expected to be fast compared to the mixing timescales.

\begin{figure}
\hspace{-1cm}
\includegraphics[width=300pt]{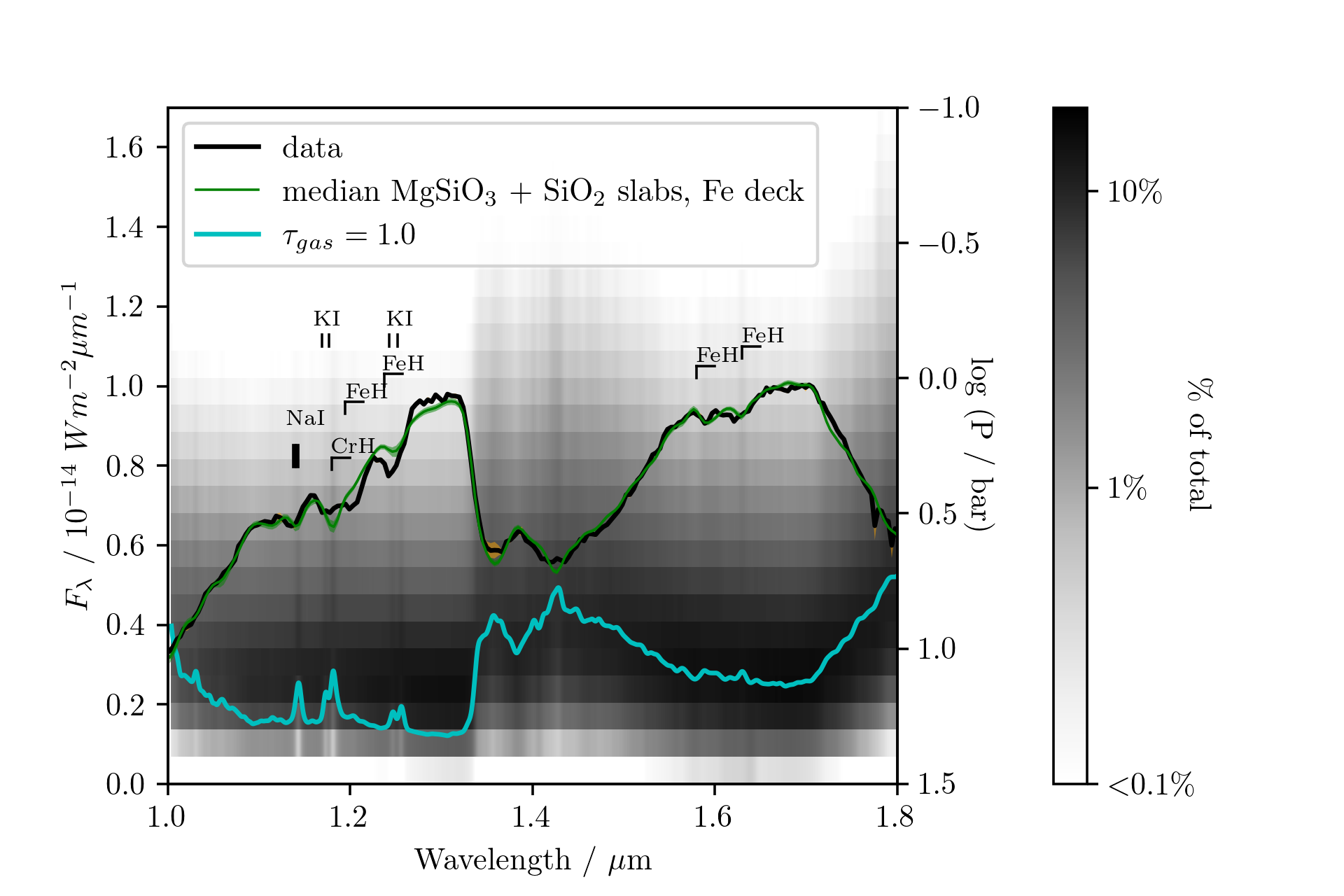}
\caption{Comparison of our preferred model $J$ and $H$ band spectrum with the data for 2M2224-0158. The $1\sigma$ spread in the model distribution is shown in green shading, and the errors for the data are shaded in orange. Both are mostly covered by the line width except in a few places. Features responsible for regions of particularly poor fit are annotated.  Also plotted, with its y-scale on the right hand side, is the contribution function. The $\tau = 1$ line for gas-phase absorption is indicated with a cyan line.
\label{fig:JHspec}}
\end{figure}

As we saw in Figure~\ref{fig:chemeq}, the equilibrium mixing fractions of several of our absorbing gases are expected to vary by orders of magnitude through the photosphere, mainly due to condensation processes. 
As a result, vertically constant mixing ratios might be expected to struggle to fit features arising from different pressure levels in the atmosphere. 
In Figure~\ref{fig:JHspec} we highlight such a case. 
The FeH features in the peak of the $H$ band are well fit by our preferred model, while the FeH features at shorter wavelengths are quite poorly fit. 
The contribution function, also shown in Figure~\ref{fig:JHspec}, demonstrates that these two regions of the spectrum are influenced by gases at different depths, with the $J$ band spectrum dominated by contributions from deeper pressures than the $H$ band spectrum.

\begin{figure}
\hspace{-1.0cm}
\includegraphics[width=300pt]{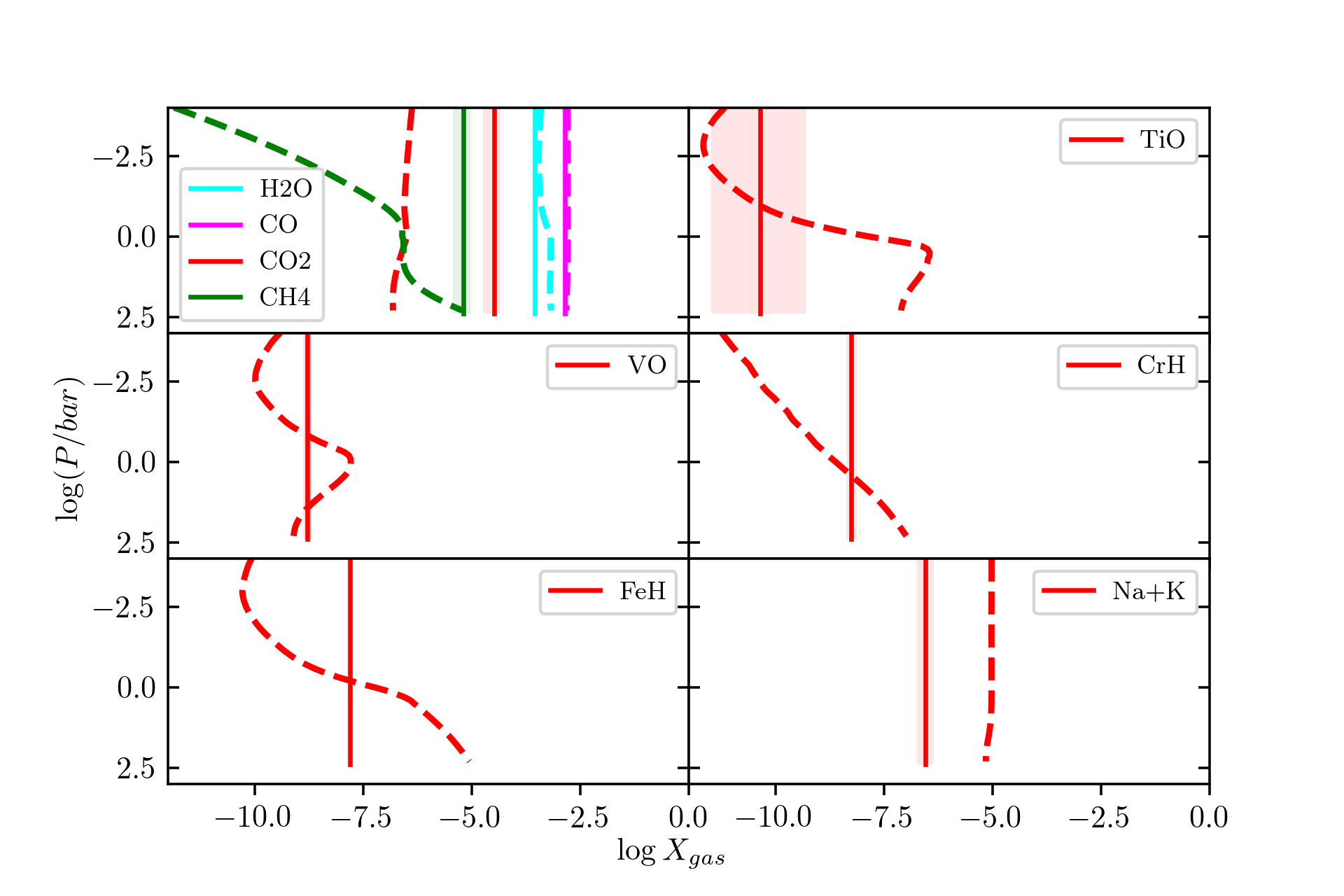}
\caption{Retrieved gas fractions compared to predictions from our thermochemical grid for our estimated metallicity and C/O ratio. 
Equilibrium predictions are shown as dashed lines and are calculated for our estimated ${\rm [M/H]}$ and C/O values. 
The solid straight lines and shading show our median retrieved values and $16^{th}$ to $84^{th}$-percentiles respectively. 
\label{fig:abund}}
\end{figure}

In Figure~\ref{fig:abund} we plot our retrieved gas mixing ratios along with predictions from our grid of thermochemical equilibrium models interpolated for our derived metallicity and C/O ratio. 
As can be seen, the equilibrium FeH fraction is expected to increase rapidly with pressure through the photosphere, so the failure of our vertically constant mixing ratios to fit both sets of features is not unexpected in this context.
We also note that our retrieved FeH fraction intersects the equilibrium prediction at a pressure just less than 1~bar. 
This is a shallower pressure than the peak in contributions for the FeH-affected spectral regions 
It follows then that the FeH abundance may be somewhat lower than predicted by the equilibrium grid at the peak contribution pressures for those features. This may contribute to rejection of the thermochemical equilibrium model.

Our model infers a $\sim 1.5$~dex lower abundance for the combined Na+K fraction than predicted by the thermochemical equilibrium grid for our extrapolated metallicity and C/O ratio. 
Our preferred model is also unable to fit the narrow \ion{K}{I} features in the near-infrared while successfully fitting the pseudo-continuum around 1$\mu$m that is set by the pressure broadened wing of the 7700\AA~\ion{K}{I} resonance line (see Figure~\ref{fig:JHspec}, despite an expected roughly vertically constant mixing ratio. 
This suggests that one reason the thermochemical equilibrium version of our model was rejected may lie in the ongoing challenge posed by correctly modelling the pressure broadened wings of the alkali lines, rather than indicating that an assumption of thermochemical equilibrium is incorrect for potassium in this type of object, or that there is some deficiency in our chemical grid.

 Our vertically constant gas fractions agree well with the predicted fractions for \hho and CO, though our retrieved estimate for H$_2$O is somewhat smaller than the equilibrium prediction deeper than around 1~bar.
 Our retrieved CO and H$_2$O fractions are also both lower by 0.5~dex compared to our estimate in B17, which used NIR data only, resulting in better agreement with thermochemical models.

By contrast, our retrieved fractions for \chhhh and \coo are both considerably higher than the equilibrium predictions. 
These non-equilibrium fractions may arise due to rapid vertical mixing, whereby CO and \chhhh are quenched at an abundance from a deeper level in the atmosphere (although CO quenches close to its equilibrium value as the dominant C-bearing gas).  
In this scenario, \coo also displays an enhanced non-equilibrium abundance due to continued reaction equilibria with CO, which increases the \coo fraction in the layers above the quench level \citep[e.g.][]{visscher2010}. 
Quenching of \chhhh at temperatures of around 1500~K at 1~bar has been been predicted \citep{visscher2011, moses2011}, but is not included in our thermochemical grid, and the implied quench temperature in this case is somewhat higher for our inferred [M/H] and C/O values. 
Although the relative abundances of \chhhh and \coo are very sensitive to [M/H], the C/O ratio \citep[e.g., see ][]{moses2013a,moses2013}, and the quench temperature, the behavior implied from the retrieved abundances suggests that rapid vertical transport may drive non-equilibrium abundances even for higher-temperature objects.

\section{Conclusions}
\label{sec:conc}

We have presented the results of a detailed retrieval analysis of the red L~dwarf 2M2224-0158.
After testing some 61 different combination of cloud opacities, we have found that the clouds appear to be dominated by layers of small grained amorphous enstatite (particle radii,  $r \lesssim 0.1~{\rm \mu m}$) and quartz ($r \sim 0.4~{\rm \mu m}$) at shallow pressures ($\lesssim 0.1~{\rm bar}$), combined with a deep iron cloud deck ($r \sim 0.1~{\rm \mu m}$) becoming optically thick at ${\rm 1~\mu m}$ at a  pressure close to 10~bar. 
The cloud opacity is best modelled by a Hansen distribution of particle sizes that scatter light according to Mie theory.
Our analysis strongly rejects the log-normal distribution and the Distribution of Hollow Spheres scattering model. 
We estimate a radius of $0.75 \pm 0.02$~\Rjup, which is considerably smaller than predicted by evolutionary models for a field age object with the luminosity of 2M2224-0158. 

Our retrieved thermal profile matches the grid model predictions as well as its flexibility allows at pressures deeper than about 1~bar.
However, we find a stratosphere that is some 500~K warmer than expected, suggestive of additional vertical heat transport not included in the self-consistent models. 

All of our highly-ranked models assume vertically constant mixing fractions for our absorbing gases, rather than those predicted by thermochemical equilibrium. This likely reflects a combination of quenched carbon chemistry, and ongoing issues with the pressure-broadened alkali opacities.

The combination of quartz and enstatite implies either that this target has a Mg/Si ratio of significantly less than 1 ($\sim 0.7$), or that phase equilibrium arguments based upon traditional assumptions (e.g. solar elemental abundance ratios) do not well predict dominant cloud opacities.
This low-value of Mg/Si is consistent with the stellar distribution, which shows significant spread. We have also estimated a high-metallicity (${\rm [M/H]} = 0.38^{+0.07}_{-0.06}$) and high-C/O ratio (${\rm 0.83^{+0.06}_{-0.07}}$), both of which lie at the upper end of the stellar distribution in the Solar Neighbourhood. 
High-metallicity is correlated with both high C/O ratios and low Mg/Si ratios in stars, so our retrieved composition appears to be telling a self-consistent story.

Predictions of cloud compositions have historically focused on solar composition atmospheres, or have assumed solar abundance ratios when scaling for metallicity. 
Since the spread in Mg/Si for exoplanets can be expected to show a similar spread to that of the stellar population \citep[e.g. ][]{bonsor2021}, our results demonstrate the need to extend both microphysical and phase equilibrium cloud models to a wider range of compositions if the full potential of upcoming JWST observations are to be realised. 
In addition, it will be necessary to test these models against cases where the ground-truth Mg/Si ratio can be estimated, such as in brown dwarf binary companions to main sequence F and G type stars.

Such a detailed treatment of a target is time consuming both in CPU hours and human effort. 
However, by exploring such a wide-range of both plausible and implausible cloud properties, we can place some confidence that our results are providing meaningful insight to the conditions in this target's atmosphere, and demonstrates our ability to test the predictions of different cloud models. 
It also highlights the potential of using cloud opacity to constrain abundance ratios such as Mg/Si. 
Future explorations using our {\it Brewster} framework will not require such a broad set of models, and extension to a larger number of LT~dwarfs is already well underway. 
Furthermore, our demonstrated ability to effectively leverage the wide ${\rm 1 - 15 \mu m}$ wavelength range makes {\it Brewster} especially well suited to exploiting JWST data for self-luminous exoplanets in the coming years.

\section*{Acknowledgements}
BB acknowledges financial support from the European Commission in the form of a Marie Curie International Outgoing Fellowship (PIOF-GA-2013- 629435). This research was made possible thanks to the Royal Society International Exchange grant No. IES/R3/170266.
E.G. acknowledges support for this work by the NSF under Grant No. AST-1614527 and Grant No. AST-1313278 and by NASA under \textit{Kepler} Grant No. 80NSSC19K0106. 
MFB and JG acknowledge support UK Research and Innovation-Science and Technology Facilities Council (UKRI-STFC) studentships. This work made use of SciPy \citep{scipy}, NumPy \citep{numpy} and matplotlib, a Python library for publication quality graphics \citep{matplotlib}, as well as Astropy, a community-developed core Python package for Astronomy \citep{astropy2013,astropy2018}.
{\it Brewster} also relies on the F2PY package \citep{f2py}.
We thank the anonymous referee for a helpful review which improved the quality of this manuscript.

\section*{Data Availability}
All data underlying this article are publicly available from the relevant
observatory archive, or upon reasonable request to the author.

\bibliographystyle{mnras}
\bibliography{refs}

\end{document}